\documentclass[
prd
,showpacs,amssymb,superscriptaddress,aps
]{revtex4}
\usepackage{graphicx}
\usepackage{color}
\input{epsf}

\usepackage{amsmath,amssymb}
\usepackage{bm}
\usepackage{times}
\usepackage{ulem}

\newcommand{\dalm}{\kern1pt\vbox{\hrule height 0.9pt\hbox{\vrule width 0.9pt
\hskip 2.5pt\vbox{\vskip 5.5pt}\hskip 3pt\vrule width 0.3pt}\hrule height 0.3pt}
\kern1pt}

\newcommand{\lsim}{\, \, \raisebox{-0.8ex}{$\stackrel{\textstyle <}{\sim}$ }}

\begin{document}



\title{Neutron star mass formula with nuclear saturation parameters}

\author{Hajime Sotani}
\email{sotani@yukawa.kyoto-u.ac.jp}
\affiliation{Astrophysical Big Bang Laboratory, RIKEN, Saitama 351-0198, Japan}
\affiliation{Interdisciplinary Theoretical \& Mathematical Science Program (iTHEMS), RIKEN, Saitama 351-0198, Japan}

\author{Hajime Togashi}
\affiliation{Department of Physics, Tohoku University, Sendai 980-8578, Japan}

\date{\today}

\begin{abstract}
We derive the empirical formulas for the neutron star mass and gravitational redshift as a function of the central density and specific combination of the nuclear saturation parameters, which are applicable to the stellar models constructed with the central density up to threefold nuclear saturation density. Combining the both empirical formulas, one also estimates the neutron star radius. In practice, we find that the neutron star mass (radius) can be estimated within $\sim 10\%$ (a few percent) accuracy by comparing the mass and radius evaluated with our empirical formulas to those determined with the specific equation of state. Since our empirical formulas directly connect the neutron star mass and radius to the nuclear saturation parameters, one can discuss the neutron star properties with the specific values of nuclear saturation parameters constrained via nuclear experiments. 
\end{abstract}

%
\maketitle


\section{Introduction}
\label{sec:I}

A neutron star is produced as a compact remnant through a supernova explosion, which occurs at the last moment of a massive star's life. The neutron stars are in extreme states, which is hard to be realized in terrestrial laboratories. In particular, due to the nature of the nuclear saturation properties, it is quite difficult to  obtain the nuclear information in a higher density region through the terrestrial experiments. This is a reason why the equation of state (EOS) for neutron star matter has not been fixed yet. Namely, the structure of the neutron star and its maximum mass are not exactly determined. Thus, the observations of the neutron stars and/or the phenomena associated with the neutron stars are quite important for understanding the physics in such extreme states.

For example, the discovery of the $2M_\odot$ neutron stars \cite{D10,A13,C20,F21} has ruled out some of the soft EOSs as the EOS for neutron star matter. 
That is, the EOS, with which the maximum mass does not reach the observed mass, can be ruled out. In addition, the light bending induced by the strong gravitational field, which is one of the important relativistic effects, modifys the pulsar light curve, which principally tells us the stellar compactness, i.e., the ratio of the stellar mass to its radius (e.g., \cite{PFC83,LL95,PG03,PO14,SM18,Sotani20a}). In practice, through the observations with the Neutron star Interior Composition Explorer (NICER) operating on an International Space Station, the mass and radius of PSR J0030+0451 \cite{Riley19,Miller19} and PSR J0740+6620 \cite{Riley21,Miller21} are constrained. Owing to the gravitational wave observations in the event of GW170817 \cite{gw170817}, the tidal deformability of the neutron star just before the merger of the binary neutron stars is also constrained, which tells us that the $1.4M_\odot$ neutron star radius should be less than $13.6$ km~\cite{Annala18}. Furthermore, it is proposed that the neutron star mass and radius may be determined with the technique of asteroseismology thorough  the future gravitational wave observations (e.g., \cite {AK1996,AK1998,STM2001,SH2003,SYMT2011,PA2012,DGKK2013,Sotani2020,SD2021}). These astronomical constraints on the neutron star mass and radius indirectly constrain the EOS for neutron star matter especially for a higher density region.

On the other hand, terrestrial experiments are obviously important for extracting the nuclear information, which also constrains the EOS for neutron star matter, even though the resultant constraint may be mainly around the nuclear saturation density. Up to now a lot of experiments worldwide have been done to fix the nuclear saturation parameters. Owing to these attempts, some of the saturation parameters have been constrained well, but many parameters, especially for higher order terms, still remain uncertain (see Sec. \ref{sec:EOS} for more detail). For instance, even the constraint on the density-dependence of the nuclear symmetry energy, which is recently reported from two large facilities in Japan and the USA, still has large uncertainties \cite{SPIRIT,PREXII}. Additionally, since the EOS for neutron star matter can be characterized by the nuclear saturation parameters, the neutron star properties may be also associated with the saturation parameters. Thus, to improve our understanding of the nuclear properties, the constraint on the neutron star mass and radius from the astronomical observations are quite important as well as constraints on the nuclear properties from the terrestrial experiments.

Nevertheless, even if one would accurately observe the neutron star mass and/or radius, it is difficult to directly discuss the nuclear saturation parameters. This is because the neutron star properties are associated with the EOSs, which can be characterized by the nuclear saturation parameters, but direct connection between the neutron star properties and nuclear saturation parameters is still unclear. To partially solve this difficulty, we have already found a suitable combination of the nuclear saturation parameters, with which the low-mass neutron star models can be expressed well \cite{SIOO14}. In this study, we extend the previous work and try to derive the empirical formulas for the mass and gravitational redshift of neutron star models constructed with the central density up to threefold nuclear saturation density, which helps us to directly discuss the association between the neutron star properties and nuclear saturation parameters.

This manuscript is organized as follows. In Sec. \ref{sec:EOS}, we briefly mention the EOSs considered in this study. In Sec. \ref{sec:NS}, we systematically examine the neutron star models and derive the empirical formulas for the neutron star mass and its gravitational redshift as a function of the nuclear saturation parameters. Finally, in Sec. \ref{sec:Conclusion}, we conclude this study. 
Unless otherwise mentioned, we adopt geometric units in the following, $c=G=1$, where $c$ and $G$ denote the speed of light and the gravitational constant, respectively.

\begin{table*}
\caption{EOS parameters adopted in this study, $K_0$, $n_0$, $L$, $Q$, $K_{sym}$, and $Q_{sym}$ are listed, while $\eta$, $\xi$, $\eta_{sy}$, and $\xi_{sy}$ are specific combinations with them given by $\eta = \left(K_0 L^2\right)^{1/3}$, $\xi = \left|{Q^6K_{sym}}/{Q_{sym}}\right|^{1/6}$, $\eta_{sy} = \left[(K_0 + K_{sym})L^2\right]^{1/3}$, and $\xi_{sy} = \left|{Q^{11}K_{sym}}/{Q_{sym}}\right|^{1/11}$. In addition, we also list the TOV mass of the neutron stars constructed with the EOSs listed here with the central density $n_c=3n_0$. } 
\label{tab:EOS2}
\begin {center}
\begin{tabular}{c|cccccc|cccc|c}
\hline\hline
EOS &  $K_0$ & $n_0$ & $L$ & $Q$  & $K_{sym}$ & $Q_{sym}$ & $\eta$ & $\xi$ &  $\eta_{sy}$ & $\xi_{sy}$ & $M_{n_c/n_0=3}$  \\
          & (MeV)& (fm$^{-3}$) & (MeV) & (MeV) &  (MeV)  &  (MeV) &  (MeV)  & (MeV) & (MeV) & (MeV) & ($M_\odot$) \\
\hline
OI-EOSs &  200 & 0.165 & 35.6  & -759 &  -142   & 801  & 63.3  & 569  & 41.8  & 649 &  0.68 \\
               &         & 0.165 & 67.8  & -761 &  -27.6  & 589  & 97.2  & 457  & 92.5  & 576 & 1.17 \\
               &  220  & 0.161 & 40.2 & -720 & -144    & 731  & 70.9  & 549  & 49.7  & 621 & 0.81 \\
               &          & 0.161 & 77.6 & -722 & -9.83   & 486  & 110  &  377  & 108   & 506 & 1.32 \\
               &  240  & 0.159 & 45.0 & -663 & -146    & 642  & 78.6 &  518  & 57.6  & 579 & 0.95 \\
               &          & 0.158 & 88.2 & -664 & 10.5    & 363  & 123  &  368  & 125   & 482 & 1.47 \\
               &  260  & 0.156 & 49.8 & -589 & -146    & 535  & 86.4 &  474  & 65.6  & 523 & 1.09 \\
               &          & 0.155 &  99.2 & -590 &  32.6  & 219  & 137 &   429  & 142   & 496 & 1.61 \\
               &  280  & 0.154 & 54.9  & -496 & -146   & 410  & 94.5 &  418  & 73.8  & 452 & 1.23 \\
               &          & 0.153 & 111   & -498 & 57.4    & 54.4 & 151  &  502  & 161   & 500 & 1.76  \\
               &  300  & 0.152 & 60.0 & -386 & -146    & 266  &  103 &  349  & 82.2  & 366 & 1.38 \\
               &          & 0.151 & 124  & -387 & 86.1    & -133 & 167  &  360  & 181   & 372 &  1.90 \\  
 KDE0v  & 229   & 0.161 & 45.2 &  -373 & -145   & 523   &  77.6 & 301  & 55.6  & 332 & 1.11 \\ 
 KDE0v1 & 228 & 0.165  & 54.7  & -385 & -127   &  484  &  88.0 & 308  & 67.0  & 341 & 1.19 \\ 
 SLy2      & 230   & 0.161 & 47.5 & -364 & -115   &  507  &  80.3 & 285  & 63.7  & 318 & 1.26 \\   
 SLy4      &  230  & 0.160 & 45.9 & -363 & -120   &  522  &  78.7 & 284  & 61.6 & 318 & 1.22 \\    
 SLy9      &  230  & 0.151 & 54.9 & -350 &  -81.4 &  462  &  88.4 & 262  & 76.4 & 299 & 1.41 \\   
 SKa       &  263  & 0.155 & 74.6 & -300 &  -78.5  & 175  &  114  &  263  & 101  & 279 & 1.57 \\    
 SkI3       & 258 & 0.158 & 101   & -304  & 73.0   &  212  & 138  &  254  & 150  & 276 & 1.77 \\   
 SkMp     & 231 & 0.157 & 70.3  & -338  & -49.8  &  159  &  105 &  278  & 96.4 & 304 & 1.45 \\  
 Shen      &  281  & 0.145  & 111    & --- &  33.5   & ---  &  151  & --- &  157  & --- & 1.82 \\  
 Togashi  &  245  & 0.160  &  38.7  & --- & --- & ---  &  71.6  & --- &  ---  & --- & 1.27 \\ 
\hline \hline
\end{tabular}
\end {center}
\end{table*}

\section{EOS for neutron star matter}
\label{sec:EOS}

\begin{figure}[tbp]
\begin{center}
\includegraphics[scale=0.5]{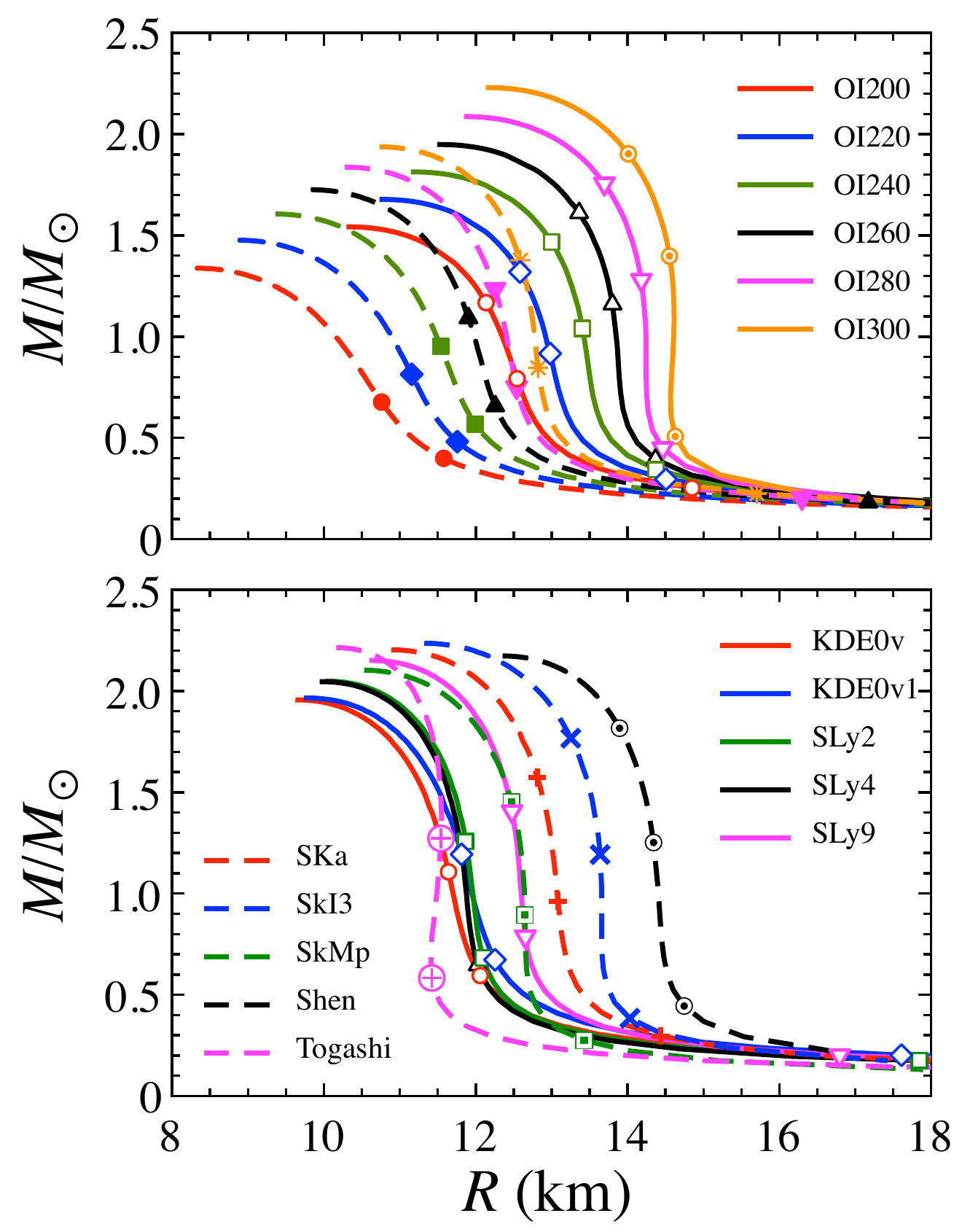} 
\end{center}
\caption{
Mass and radius relation for the neutron star models constructed with the EOSs listed in Table \ref{tab:EOS2}, where the top and bottom panels correspond to the result with OI-EOSs and the others, respectively. The OI-EOSs are named with the value of $K_0$, e.g., OI200 for the OI-EOSs with $K_0=200$ MeV. In the top panel, the solid and dashed lines respectively denote the OI-EOSs with larger and smaller values of $L$ for each value of $K_0$ (see Table \ref{tab:EOS2}). In each panel, the neutron star models with $n_c/n_0=1$, 2, and 3 are shown with the marks.
}
\label{fig:MR}
\end{figure}

To construct the neutron star models by solving the Tolman-Oppenheimer-Volkoff (TOV) equation, one has to assume an EOS for neutron star matter. In this study, we mainly adopt
the phenomenological nuclear EOS models, focusing only on the unified EOS, i.e., the neutron star crust EOS is constructed with the same nuclear model as in the neutron star core EOS. As a phenomenological macroscopic model, we adopt the EOSs proposed by Oyamatsu and Iida (hereafter referred to as the OI-EOS) \cite{OI03,OI07}. The OI-EOSs are constructed with the Pad\'{e}-type potential energies in such a way as to reproduce empirical masses and radii of stable nuclei, using a simplified version of the extended Thomas-Fermi theory. 
On the other hand, as a phenomenological Skyrme-type model, we adopt KDE0v, KDE0v1 \cite{KDE0v}, SLy2, SLy4, SLy9 \cite{SLy4,SLy9}, SKa \cite{SKa}, SkI3 \cite{SkI3}, and SkMp \cite{SkMp}. In addition, we also adopt the Shen EOS \cite{Shen}, which is based on the relativistic mean field theory, and the Togashi EOS \cite{Togashi17}, which is derived by the variational many-body calculation with AV18 two-body and UIX three-body potentials.
In Fig. \ref{fig:MR}, we show the mass and radius relation for the neutron star models constructed with the EOSs adopted in this study, where the stellar models with $n_c/n_0=1$, 2, and 3 are shown with the marks. We note that some of the stellar models with $n_c/n_0=1$ are out of the panel due to the large radius. One can observe that some of EOSs are obviously ruled out from the $2M_\odot$ observations \cite{D10,A13,C20,F21} or the radius constraint from the GW170817 \cite{Annala18}, but in order to examine with the wide parameter space, we adopt even such EOSs in this study.

In any case, the bulk energy per nucleon for the uniform nuclear matter at zero temperature can generally be expressed as a function of the baryon number density, $n_{\rm b}=n_n+n_p$, and an asymmetry parameter, $\alpha=(n_n-n_p)/n_{\rm b}$, with the neutron number density, $n_n$, and the proton number density, $n_p$; 
\begin{equation}
  \frac{E}{A} = w_s(n_{\rm b}) + \alpha^2 S(n_{\rm b}) + {\cal O}(\alpha^3), \label{eq:E/A}
\end{equation}
where $w_s$ corresponds to the energy per nucleon of symmetric nuclear matter, while $S$ denotes the density-dependent symmetry energy. Additionally, $w_s$ and $S$ can be expanded around the saturation density, $n_0$, of the symmetric nuclear matter as a function of $u=(n_{\rm b}-n_0)/(3n_0)$;
\begin{gather}
  w_s(n_{\rm b}) = w_0 + \frac{K_0}{2}u^2 + \frac{Q}{6}u^3 + {\cal O}(u^4), \label{eq:ws} \\
  S(n_{\rm b}) = S_0 + Lu + \frac{K_{sym}}{2}u^2 + \frac{Q_{sym}}{6}u^3 + {\cal O}(u^4). \label{eq:S}
\end{gather}
The coefficients in these expressions are the nuclear saturation parameters, with which each EOS is characterized. The parameters for the adopted EOSs are concretely listed in Table \ref{tab:EOS2}, where $\eta$, $\xi$, $\eta_{sy}$, and $\xi_{sy}$ are the specific combination of the nuclear saturation parameters (see the following sections for details), defined by 
\begin{gather}
  \eta = \left(K_0 L^2\right)^{1/3}, \label{eq:eta} \\
  \xi = \left|\frac{Q^6K_{sym}}{Q_{sym}}\right|^{1/6}, \label{eq:xi} \\
  \eta_{sy} = \left[(K_0 + K_{sym})L^2\right]^{1/3}, \label{eq:etasy} \\
  \xi_{sy} = \left|\frac{Q^{11}K_{sym}}{Q_{sym}}\right|^{1/11}.  \label{eq:xisy} 
\end{gather}

Among the nuclear saturation parameters, $n_0$, $w_0$, and $S_0$ are well constrained as $n_0\approx 0.15-0.16$ fm$^{-3}$, $w_0 \approx -15.8$~MeV \cite{OHKT17}, and $S_0 \approx 31.6\pm 2.7$ MeV \cite{Li19}. Meanwhile, $K_0$ and $L$ are more difficult to be determined from the terrestrial experiments, because these parameters are the density derivative at the saturation point, i.e., one needs to know the information not only at the saturation point but also in wider range around the saturation point. The constraints on these parameters are gradually improved and the current fiducial values are $K_0=230 \pm 40$ MeV \cite{KM13} and $L=58.9 \pm 16$ MeV \cite{Li19}, even though the constraints on $L$ recently reported from two large facilities still have a large uncertainty, i.e., $42\le L\le 117$ MeV with S$\pi$RIT by the Radioactive Isotope Beam Factory at RIKEN in Japan \cite{SPIRIT} and $L = 106 \pm 37$ MeV with PREX-II by the Thomas Jefferson National Accelerator Facility in Newport News, the United States \cite{PREXII}. Moreover, the saturation parameters in higher order terms, such as $Q$, $K_{sym}$, $Q_{sym}$, are almost unconstrained from the experiments, but they are theoretically predicted as $-800 \le Q \le 400$ MeV, $-400 \le K_{sym} \le 100$ MeV, and $-200 \le Q_{sym} \le 800$ MeV \cite{Li19}.

\begin{figure}[tbp]
\begin{center}
\includegraphics[scale=0.5]{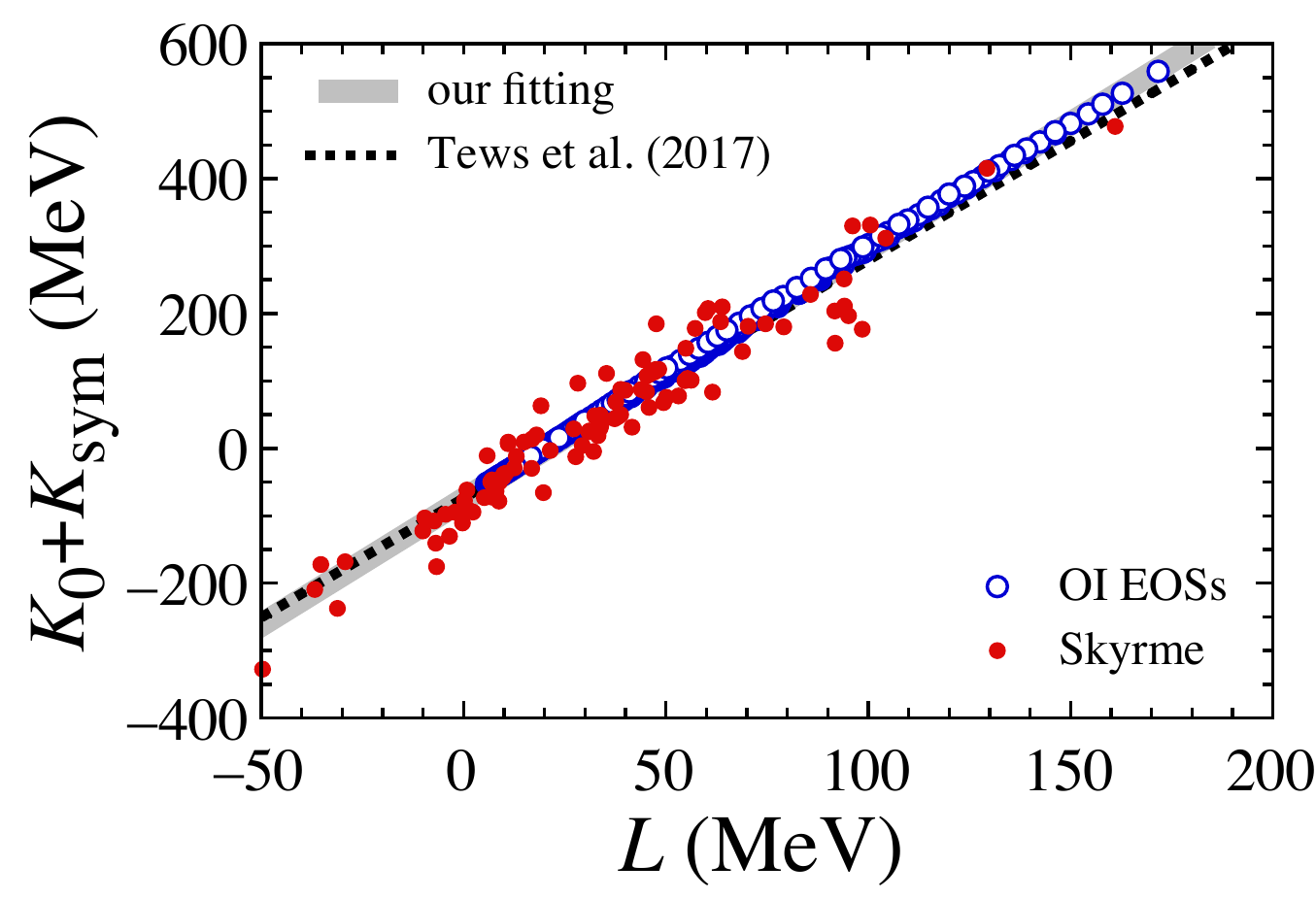} 
\end{center}
\caption{
The relation between $K_0+K_{sym}$ and $L$ for the OI-EOSs and Skyrme-type EOSs. The solid line denotes the fitting given by Eq. (\ref{eq:K0KsyL}), while the dotted line denote the fitting proposed in the previous study \cite{Tews17}.
}
\label{fig:L0Ks-L}
\end{figure}

It is known that $S_0$ is strongly associated with $L$ as $S_0\approx 28 + 0.075L$ \cite{OI03,LL13}. In a similar way, we find that $K_0+K_{sym}$ is also strongly associated with $L$, adopting 118 models for the Skyrme-type EOSs listed in Ref. \cite{Danielewicz09} and 304 models for OI-EOSs. We plot $K_0+K_{sym}$ as a function of $L$ in Fig. \ref{fig:L0Ks-L}, where the thick solid line denotes the fitting formula given by
\begin{equation}
  K_0+K_{sym} = -75.86 + 371.8 \left(\frac{L}{100\ {\rm MeV}}\right).   \label{eq:K0KsyL}
\end{equation}
The similar correlation has been reported in Ref. \cite{Tews17}, which is shown in Fig. \ref{fig:L0Ks-L} with the dotted line. This type of correlation may be  very useful for constraining the value of $K_{sym}$ with using the constraints on $K_0$ and $L$, because the uncertainty in $K_{sym}$ is still very large. In practice, by assigning the fiducial values of $K_0$ and $L$ mentioned above in Eq. (\ref{eq:K0KsyL}), one can find that $-186\le K_{sym} \le 13$ MeV.

\section{Neutron star mass formula}
\label{sec:NS}

The neutron star structure is determined by solving the TOV equations together with the appropriate EOS. The neutron star may sometimes be considered as a huge nucleus, 
but its structure is quite dense, compared to atomic nuclei. Nevertheless, since the density inside low-mass neutron stars is definitely low, their mass 
seems to be strongly associated with the nuclear saturation properties. In practice, it has been found that the mass, $M$, and gravitational redshift, $z\equiv(1-2GM/Rc^2)^{-1/2}-1$, for the low-mass neutron stars, whose central density is less than twice the nuclear saturation density, are well expressed as a function of $\eta$ defined by Eq. (\ref{eq:eta})
and $u_c=\rho_c/\rho_0$, where $\rho_c$ and $\rho_0$ are the central energy density and the energy density  corresponding to the nuclear saturation density, i.e., $\rho_0=2.68\times 10^{14}$ g/cm$^3$ \cite{SIOO14}. That is, one can estimate neutron star mass and radius by combining the empirical formulas, $M=M(u_c,\eta)$ and $z=z(u_c,\eta)$. In practice, assuming the recent experimental constraints obtained with S$\pi$RIT and PREX-II \cite{SPIRIT,PREXII}, one can show the allowed region in the neutron star mass and radius relation \cite{SNN22}. Using this new parameter $\eta$, one can also discuss the rotational properties of the low-mass neutron stars \cite{SSB16} and the possible maximum mass of neutron stars \cite{Sotani17,SK17}.  In this study, we try to extend this type of empirical formulas even for higher central density up to three times saturation density. 
This is because, as the central density becomes larger, the empirical formulas discussed in more detail below lose accuracy. This may come from the additional EOS dependence, such as higher order coefficients in Eqs. (\ref{eq:ws}) and (\ref{eq:S}).  
In addition, for reference, we show the mass of neutron stars with $n_c=3n_0$ constructed with the EOSs considered in this study in Table \ref{tab:EOS2}, with which one may be able to adopt our empirical relations discussed below if the stellar mass is less than $\simeq 0.68 - 1.90M/M_\odot$ (average value is $1.34M_\odot$).

\subsection{Function of $\eta$}
\label{sec:NS1}

\begin{figure}[tbp]
\begin{center}
\includegraphics[scale=0.5]{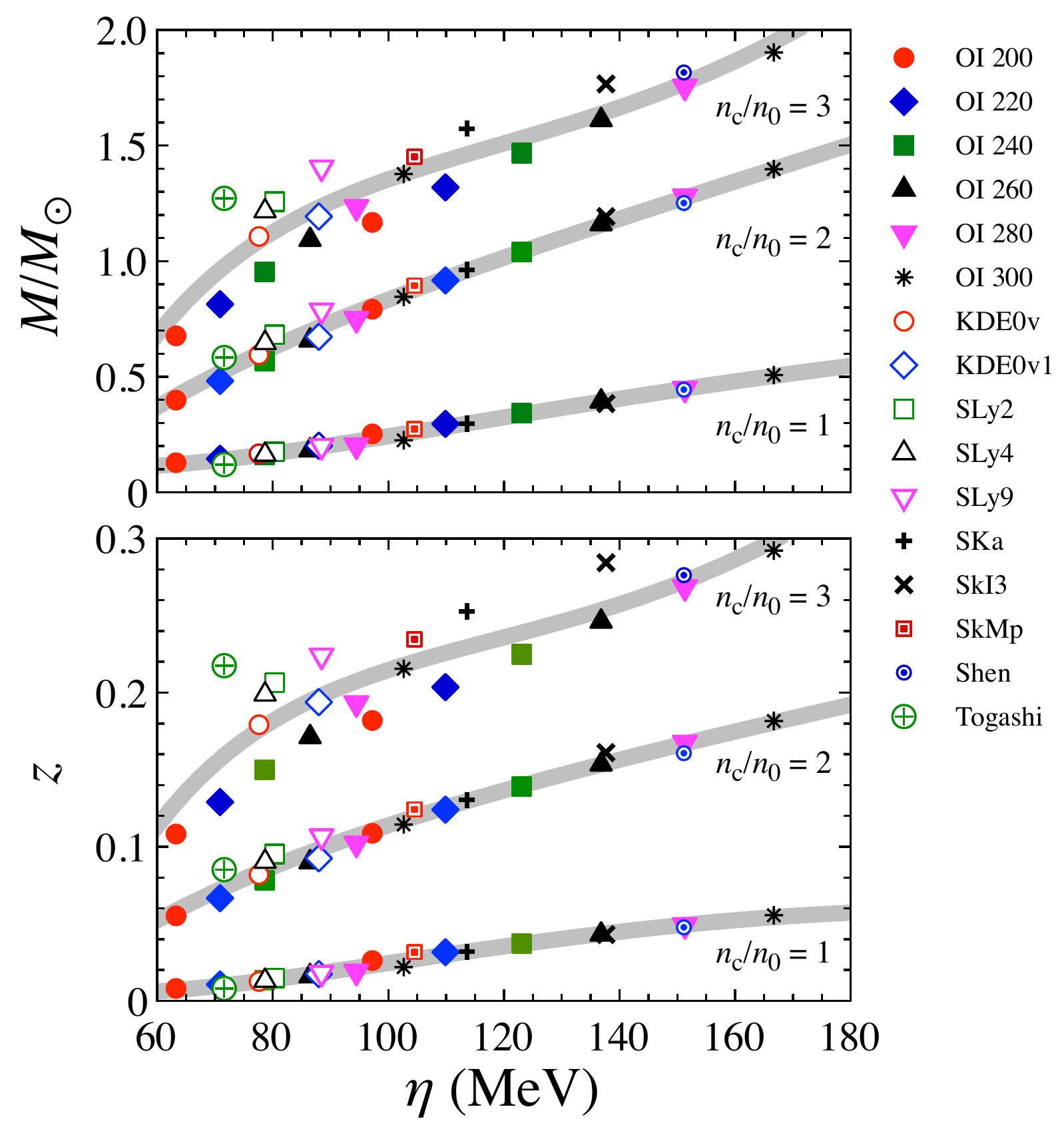} 
\end{center}
\caption{
The neutron star mass, $M$, and gravitational redshift, $z$, for various EOSs with the fixed central baryon number density, $n_c/n_0=1,2,3$, are plotted as a function of $\eta$. This thick-solid lines denote the fitting lines given by Eqs. (\ref{eq:mm}) and (\ref{eq:zz}).
}
\label{fig:Mzeta}
\end{figure}

\begin{figure}[tbp]
\begin{center}
\includegraphics[scale=0.5]{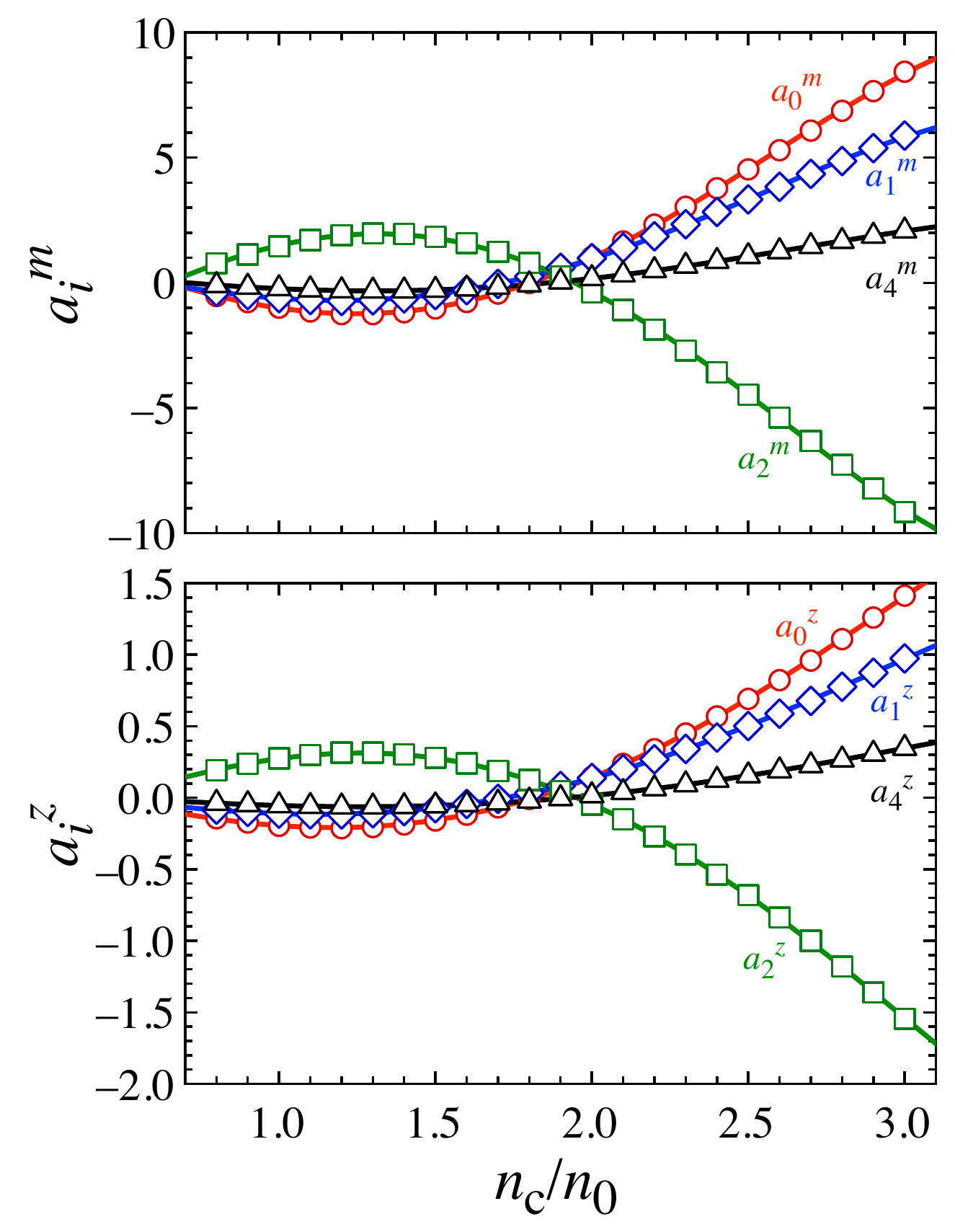} 
\end{center}
\caption{
The $n_c/n_0$ dependence of the coefficients, $a_i^m$ and $a_i^z$, in the fitting formula of stellar mass and gravitational redshift as a function of $\eta/(100\ {\rm MeV})$ expressed by Eqs. (\ref{eq:mm}) and (\ref{eq:zz}). The open marks denote the values of $a_i^m$ and $a_i^z$ in the fitting formulas, while the solid lines denote the fitting as a function of $n_c/n_0$ as Eqs. (\ref{eq:am}) and (\ref{eq:az}).
}
\label{fig:aimz}
\end{figure}

Since the saturation density, $n_0$, also depends on the EOS models, as shown in Table \ref{tab:EOS2}, it may be better to consider the mass and redshift for the low-mass neutron star as a function of $n_c/n_0$ instead of $\rho_c/\rho_0$ with the fixed value of $\rho_0$, where $n_c$ is the baryon number density at the stellar center. In fact, as shown in Fig. \ref{fig:Mzeta}, one can observe that the neutron star mass, $M$, and gravitational redshift, $z$, with the fixed central baryon number density, e.g., $n_c/n_0=1,2,3$, are strongly correlated with $\eta$. We note that $M$ and $z$ are quite similar dependence on $\eta$, even though each value is completely different, as shown in Ref. \cite{SIOO14}. With this result, we can derive the fitting formulas as
\begin{gather}
  \frac{M_{\eta}}{M_\odot} = a_{0}^m + a_{1}^m\ln(\eta_{100}) + a_{2}^m\eta_{100} + a_{3}^m\eta_{100}^2, 
      \label{eq:mm} \\
  z_{\eta} = a_{0}^z + a_{1}^z\ln(\eta_{100}) + a_{2}^z\eta_{100} + a_{3}^z\eta_{100}^2, \label{eq:zz} 
\end{gather}
where $\eta_{100}\equiv \eta/(100\ {\rm MeV})$, while $a_{i}^m$ and $a_{i}^z$ for $i=0-3$ are the coefficients in the fitting formulas, depending on the normalized central density, ${\cal R}_c\equiv n_c/n_0$. Here, in order to distinguish the mass and radius determined by integrating TOV equations with each EOS, the mass and redshift estimated with the fitting formulas given by Eqs. (\ref{eq:mm}) and (\ref{eq:zz}) are referred to as $M_\eta$ and $z_\eta$. In addition, as shown in Fig. \ref{fig:aimz}, we find that the coefficients $a_i^m$ and $a_i^z$ are well expressed as a function of ${\cal R}_c$ as
\begin{gather}
  a_i^m({\cal R}_c) = a_{i0}^m +a_{i1}^m{\cal R}_c +a_{i2}^m{\cal R}_c^3 +a_{i3}^m {\cal R}_c^5, \label{eq:am} \\
  a_i^z({\cal R}_c) = a_{i0}^z +a_{i1}^z{\cal R}_c +a_{i2}^z{\cal R}_c^3 +a_{i3}^z {\cal R}_c^5, \label{eq:az} 
\end{gather}
where the exact values of $a_{ij}^m$ and $a_{ij}^z$ for $i=0-3$ and $j=0-3$ are listed in Table \ref{tab:coefficients}. Now, we can get the empirical formulas for the neutron star mass and redshift as $M_\eta=M_\eta({\cal R}_c,\eta)$ and $z_\eta=z_\eta({\cal R}_c,\eta)$ given by Eqs. (\ref{eq:mm}) -- (\ref{eq:az}).

\begin{table}
\caption{Values of $a_{ij}^m$ and $a_{ij}^z$ in Eqs. (\ref{eq:am}) and (\ref{eq:az}).} 
\label{tab:coefficients}
\begin {center}
\begin{tabular}{ccccc}
\hline\hline
 $j$  &  0 & 1 & 2  & 3     \\
\hline
 $a_{0j}^m$ &  $3.2110$   & $-5.5024$  & $1.3375$  & $-0.059490$   \\
 $a_{1j}^m$ &  $1.8340$   & $-3.2893$  & $0.87206$  & $-0.039859$   \\
 $a_{2j}^m$ &  $-4.3905$  & $7.4582$  & $-1.6450$  & $0.071406$   \\
 $a_{3j}^m$ &  $0.94544$  & $-1.5004$  & $0.33522$  & $-0.014110$   \\
 $a_{0j}^z$  &  $0.25385$  & $-0.60188$  & $0.15808$  & $-0.0053861$   \\
 $a_{1j}^z$  &  $0.15141$  & $-0.36133$  & $0.10275$  & $-0.0035881$   \\
 $a_{2j}^z$  &  $-0.38040$  & $0.84306$  & $-0.19442$  & $0.0064246$   \\
 $a_{3j}^z$  &  $0.086653$  & $-0.18150$  & $0.041746$  & $-0.0013309$   \\
\hline \hline
\end{tabular}
\end {center}
\end{table}

\begin{figure}[tbp]
\begin{center}
\includegraphics[scale=0.5]{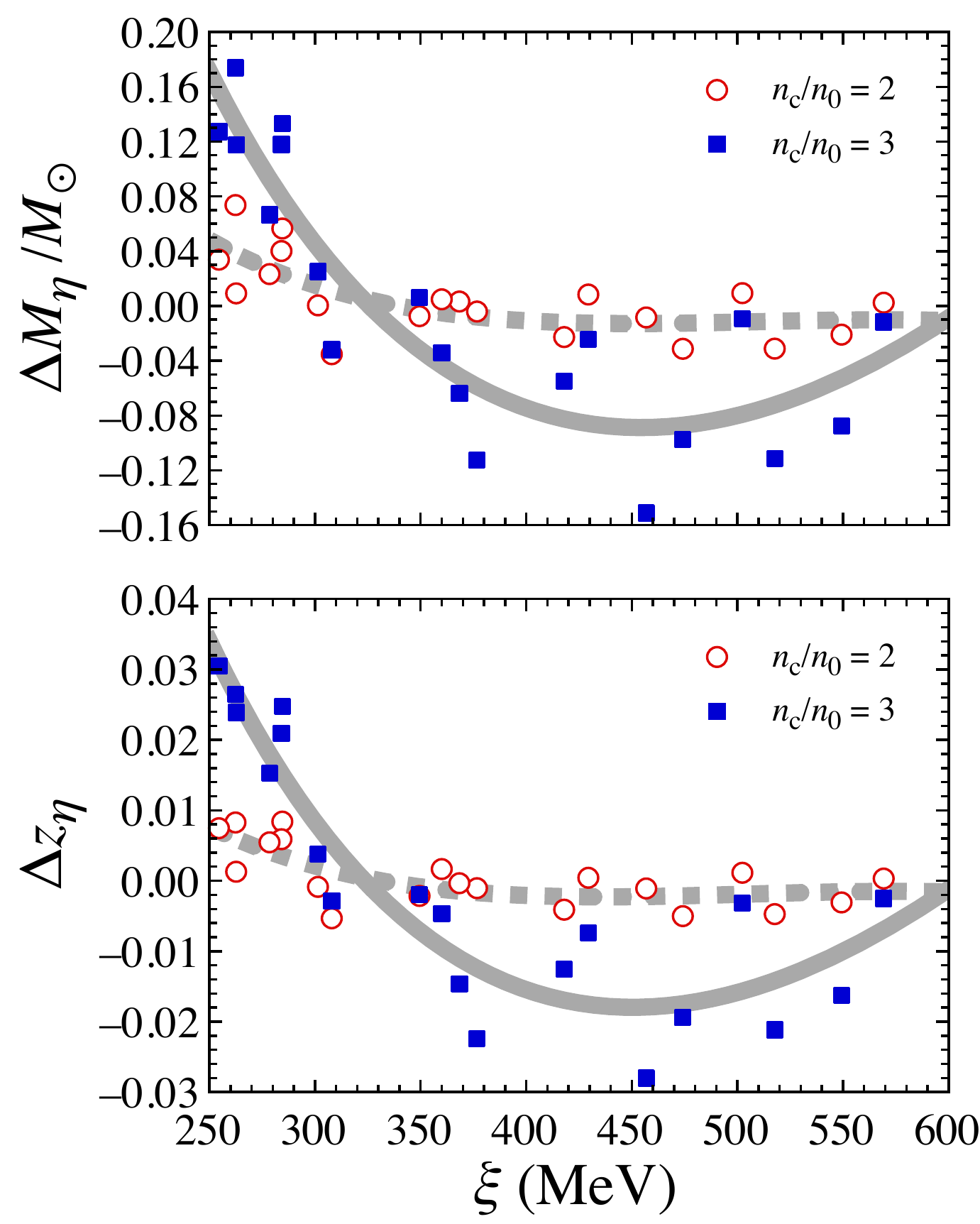} 
\end{center}
\caption{
Considering the OI-EOSs and the Skyrme-type EOSs, $\Delta M_\eta/M_\odot$ and $z_\eta$ calculated with Eqs. (\ref{dM}) and (\ref{dz}) are shown as a function of $\xi$ for $n_c/n_0 = 2$ (open circles) and 3 (filled squares). The dotted and solid lines are fitting lines given by Eqs. (\ref{eq:dMxi}) and (\ref{eq:dzxi}).
}
\label{fig:dMdz-xi}
\end{figure}

\begin{figure}[tbp]
\begin{center}
\includegraphics[scale=0.5]{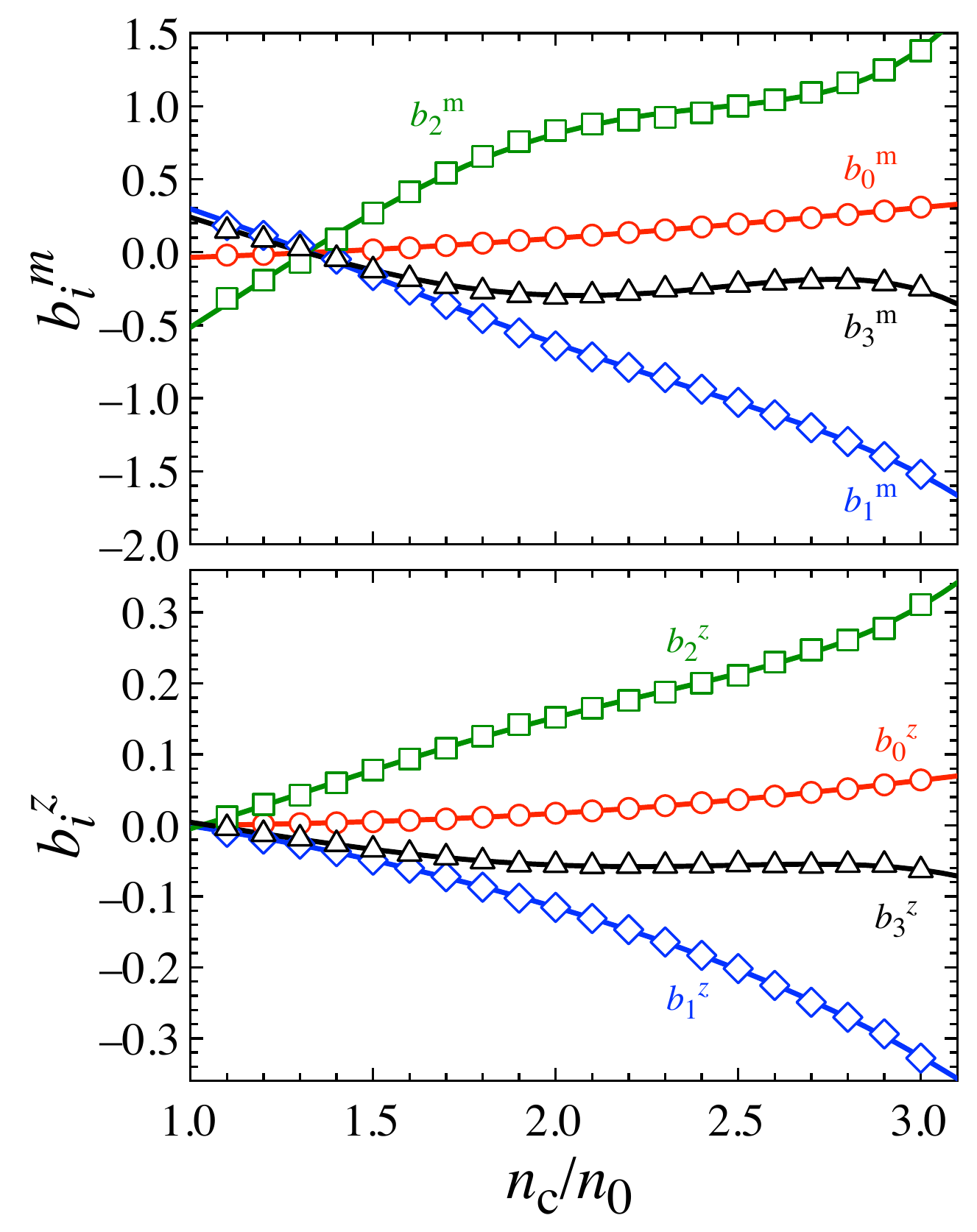} 
\end{center}
\caption{
The coefficients in Eqs. (\ref{eq:dMxi}) and (\ref{eq:dzxi}) are shown as a function of $n_c/n_0$ with marks, while the solid lines denote the fitting given by Eqs. (\ref{eq:bm}) and (\ref{eq:bz}), respectively. 
}
\label{fig:bimz}
\end{figure}

Next, in order to improve the resultant empirical formulas, we try to characterize the deviation of the neutron star mass and redshift determined with each EOS from those estimated with the fitting formulas given by Eqs. (\ref{eq:mm}) -- (\ref{eq:az}), using a specific combination of the nuclear saturation parameters in the higher order terms. That is, the deviation of the mass and redshift are given by 
\begin{gather}
  \Delta M_{\eta} = M_{\rm TOV} - M_{\eta}({\cal R}_c,\eta), \label{dM} \\
  \Delta z_{\eta} = z_{\rm TOV} - z_{\eta}({\cal R}_c,\eta), \label{dz}   
\end{gather}
where $M_{\rm TOV}$ and $z_{\rm TOV}$ are the neutron star mass and redshift determined by integrating the TOV equations together with each EOS. Through a trial and error process, we find a good combination of $K_{sym}$, $Q$, and $Q_{sym}$, which is $\xi$ defined by Eq. (\ref{eq:xi}), for characterizing $\Delta M_{\eta}$ and $\Delta z_\eta$, even though it may be not so tight correlation.
We note that it may be necessary to modify the definition of $\xi$, if $\xi$ is out of the range considered in this study, i.e.,  $250\lsim \xi \lsim 600$ MeV with the EOSs adopted in this study. In fact, $\xi$ is not defined when $Q_{sym}=0$. In practice, for the neutron star models with $n_c/n_0=2$ and 3, we show $\Delta M_\eta/M_\odot$ and $z_\eta$ as a function of $\xi$ in Fig. \ref{fig:dMdz-xi}, considering the OI-EOSs and the Skyrme-type EOSs. In this figure, we also plot the fitting formulas given by 
\begin{gather}
  \frac{\Delta M_{\eta}}{M_\odot} = b_0^m/\xi_{500}+b_1^m \xi_{500} + b_2^m \xi_{500}^2 + b_3^m \xi_{500}^3, 
       \label{eq:dMxi}  \\
  \Delta z_{\eta} = b_0^z/\xi_{500}+b_1^z \xi_{500} + b_2^z \xi_{500}^2 + b_3^z \xi_{500}^3,  \label{eq:dzxi} 
\end{gather}
where $\xi_{500}\equiv \xi/(500\ {\rm MeV})$, while again $b_i^m$ and $b_i^z$ for $i=0-3$ are the adjusting coefficients depending on ${\cal R}_c$. In a similar way for deriving the fitting formulas for $\Delta M_\eta$ and $\Delta z_\eta$, the coefficients in Eqs. (\ref{eq:dMxi}) and (\ref{eq:dzxi}) are plotted in Fig. \ref{fig:bimz} as a function of $n_c/n_0$, which can be fitted with
\begin{gather}
  b_i^m({\cal R}_c) = b_{i0}^m +b_{i1}^m{\cal R}_c^2 +b_{i2}^m{\cal R}_c^4 +b_{i3}^m {\cal R}_c^6, \label{eq:bm} \\
  b_i^z({\cal R}_c) = b_{i0}^z +b_{i1}^z{\cal R}_c^2 +b_{i2}^z{\cal R}_c^4 +b_{i3}^z {\cal R}_c^6, \label{eq:bz}
\end{gather}
where the concrete values of the coefficients $b_{ij}^m$ and $b_{ij}^z$ for $i=0-3$ and $j=0-3$ are listed in Table \ref{tab:bm}. We note that we consider the fitting of $\Delta M_\eta$ and $\Delta z_\eta$ only for $n_c/n_0 > 1$ as in Fig. \ref{fig:bimz}, even though in principle one can also fit them for lower density region. This is because the correlation of $\Delta M_\eta$ and $\Delta z_\eta$ with $\xi$ becomes weaker and the absolute values of $\Delta M_\eta$ and $\Delta z_\eta$ become much smaller, as the density becomes lower. So, for $n_c/n_0\le 1$ we simply assume that $\Delta M_\eta=0$ and $\Delta z_\eta=0$ in this study.

\begin{table*}
\caption{Values of $b_{ij}^m$ and $b_{ij}^z$ in Eqs. (\ref{eq:bm}) and (\ref{eq:bz}).} 
\label{tab:bm}
\begin {center}
\begin{tabular}{ccccc}
\hline\hline
 $j$  &  0 & 1 & 2  & 3     \\
\hline
 $b_{0j}^m$ &  $-7.711\times 10^{-2}$  & $4.122 \times 10^{-2}$  & $8.051\times 10^{-4}$  & $-7.267\times 10^{-5}$   \\
 $b_{1j}^m$ &  $7.763\times 10^{-1}$  & $-5.341\times 10^{-1}$  & $5.793\times 10^{-2}$  & $-2.995\times 10^{-3}$   \\
 $b_{2j}^m$ &  $-1.470$  & $1.123$  & $-1.774\times 10^{-1}$  & $9.771\times 10^{-3}$   \\
 $b_{3j}^m$ &  $7.276\times 10^{-1}$  & $-5.897\times 10^{-1}$  & $1.077\times 10^{-1}$  & $-6.041\times 10^{-3}$   \\
 $b_{0j}^z$  &  $-2.783\times 10^{-3}$  & $1.679\times 10^{-3}$  & $9.658\times 10^{-4}$  & $-3.707\times 10^{-5}$   \\
 $b_{1j}^z$  &  $3.989\times 10^{-2}$  & $-4.167\times 10^{-2}$  & $1.204\times 10^{-3}$  & $-1.209\times 10^{-4}$   \\
 $b_{2j}^z$  &  $-9.514\times 10^{-2}$  & $1.031\times 10^{-1}$  & $-1.341\times 10^{-2}$  & $7.713\times 10^{-4}$   \\
 $b_{3j}^z$  &  $5.321\times 10^{-2}$  & $-5.814\times 10^{-2}$  & $9.859\times 10^{-3}$  & $-5.368\times 10^{-4}$   \\
\hline \hline
\end{tabular}
\end {center}
\end{table*}

\begin{figure*}[tbp]
\begin{center}
\includegraphics[scale=0.5]{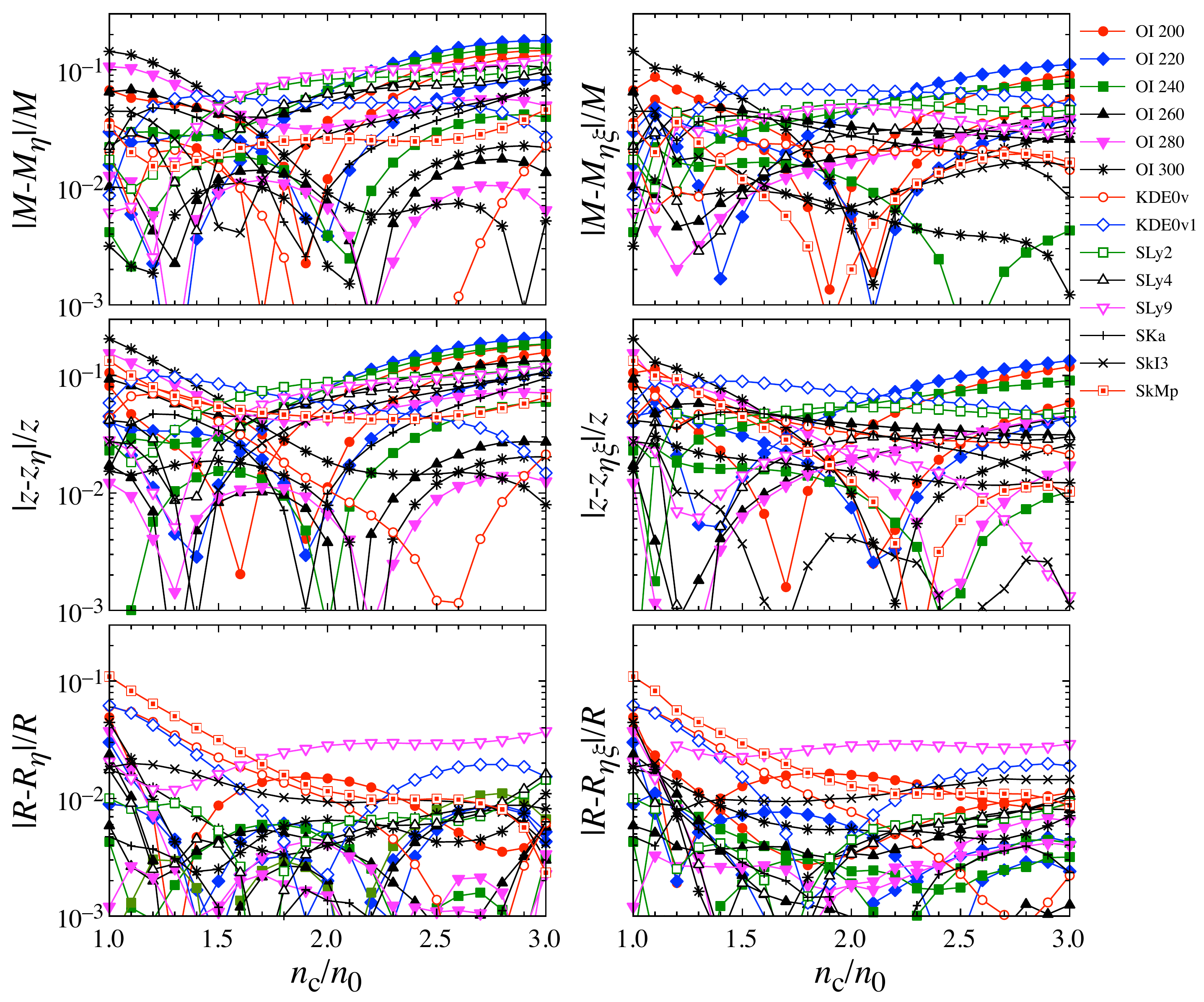} 
\end{center}
\caption{
Relative deviation of the neutron star mass (top panel) and redshift (middle panel) estimated with the empirical formulas from those constructed with each EOS are shown as a function of the normalized central baryon number density. The left panels are the deviation with the empirical formulas, $M_\eta({\cal R}_c,\eta)$ and $z_\eta({\cal R}_c,\eta)$, while the right panels are that with $M_{\eta\xi}({\cal R}_c,\eta,\xi)$ and $z_{\eta\xi}({\cal R}_c,\eta,\xi)$. The bottom panels are the relative deviation of the neutron star radius estimated with the empirical formulas for the mass and redshift from that calculated with each EOS. 
}
\label{fig:dMzR}
\end{figure*}

Now, we can derive new empirical formulas for the neutron star mass, $M_{\eta\xi}$, and redshift, $z_{\eta\xi}$, as a function of ${\cal R}_c (=n_c/n_0)$, $\eta$, and $\xi$:
\begin{gather}
  \frac{M_{\eta\xi}}{M_\odot} = \frac{M_\eta({\cal R}_c,\eta)}{M_\odot} + \frac{\Delta M_\eta({\cal R}_c,\xi)}{M_\odot}, 
     \label{eq:Mxi} \\ 
  z_{\eta\xi} = z_{\eta}({\cal R}_c,\eta) + \Delta z_\eta({\cal R}_c,\xi), \label{eq:zxi}
\end{gather}
where the first terms are given by Eqs. (\ref{eq:mm}) -- (\ref{eq:az}) and the second terms are given by Eqs. (\ref{eq:dMxi}) -- (\ref{eq:bz}). In order to check the accuracy of our empirical formulas, $M_\eta({\cal R}_c,\eta)$, $z_\eta({\cal R}_c,\eta$), $M_{\eta\xi}({\cal R}_c,\eta,\xi)$, and $z_{\eta\xi}({\cal R}_c,\eta,\xi)$, in Fig. \ref{fig:dMzR} we show the relative deviation from the neutron star mass and redshift determined through the TOV equations, where the bottom panels are the relative deviation of the neutron star radius estimated with the empirical formulas for the mass and redshift from the TOV solution. From this figure, one can see that the neutron star mass is estimated within $\sim 10\%$ accuracy, while the radius for the canonical neutron star is estimated within $\sim 3\%$ accuracy, using the empirical formulas, $M_{\eta\xi}({\cal R}_c,\eta,\xi)$ and $z_{\eta\xi}({\cal R}_c,\eta,\xi)$. We also make a comment that the mass estimation with $M_\eta({\cal R}_c,\eta)$ (top left panel) is better than that with $M_{\eta\xi}({\cal R}_c,\eta,\xi)$ (top right panel) in the density region around $n_c/n_0\simeq 1.5$, which comes from the fact that the correlation between $\Delta M_\eta$ and $\xi$ becomes worse as the density becomes lower. 
We note that the mass and gravitational redshift have quite similar dependence on $\eta$, as shown in Fig. \ref{fig:Mzeta}, i.e., it seems to get a small amount of difference in information from the mass and gravitational redshift. Even so, one can accurately recover the radius, using the empirical relations for the mass and gravitational redshift.

At the end, we mention another possibility for characterizing $\Delta M_\eta$ and $\Delta z_\eta$ instead of $\xi$. In practice, we find that a new parameter, $\xi_{{\cal R}_c}$, defined by
\begin{equation}
  \xi_{{\cal R}_c} = \left|\frac{Q^{2{\cal R}_c}K_{sym}}{Q_{sym}}\right|^{1/2{\cal R}_c}  \label{eq:xiRc} 
\end{equation}
seems to be better than $\xi$ given by Eq. (\ref{eq:xi}) for characterizing $\Delta M_\eta$ and $\Delta z_\eta$. That is,  adopting the same functional form as Eqs. (\ref{eq:dMxi}) and (\ref{eq:dzxi}), one can express $\Delta M_\eta$ and $\Delta z_\eta$ with  $\xi_{{\cal R}_c}$, where the correlation in a lower density region is better than the case with $\xi$. But, unfortunately, the dependence of the coefficients $b_i^m$ and $b_i^z$ on ${\cal R}_c$ becomes more complex. So, in this study, we simply adopt $\xi$ as mentioned above.

\subsection{Function of $\eta_{sy}$}
\label{sec:NS2}

\begin{figure}[tbp]
\begin{center}
\includegraphics[scale=0.5]{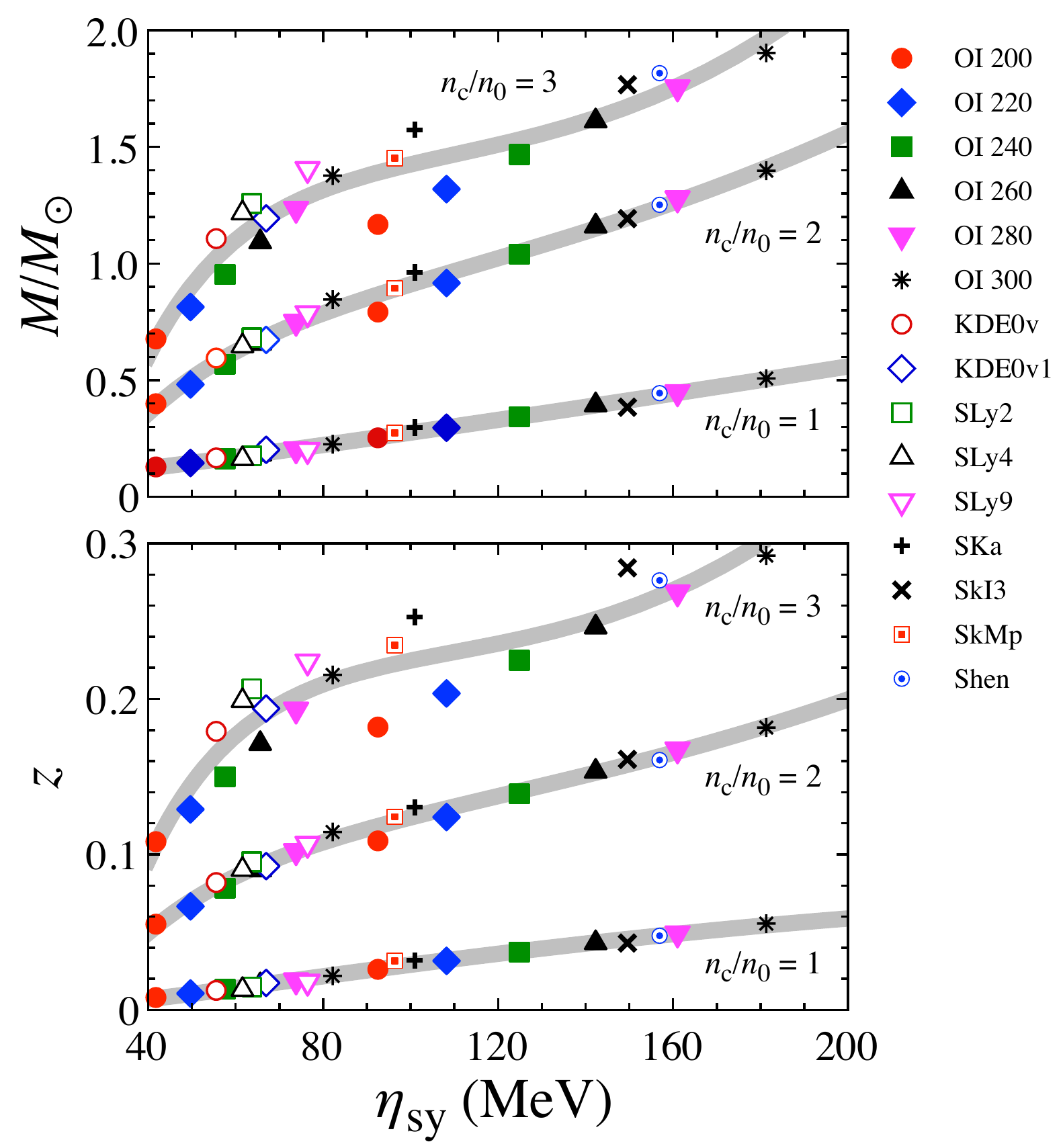} 
\end{center}
\caption{
Same as in Fig. \ref{fig:Mzeta}, but as a function of $\eta_{sy}$. The thick-solid lines denote the fitting lines given by Eqs. (\ref{eq:mmsy}) and (\ref{eq:zzsy}).
}
\label{fig:Mzetasy}
\end{figure}

\begin{figure}[tbp]
\begin{center}
\includegraphics[scale=0.5]{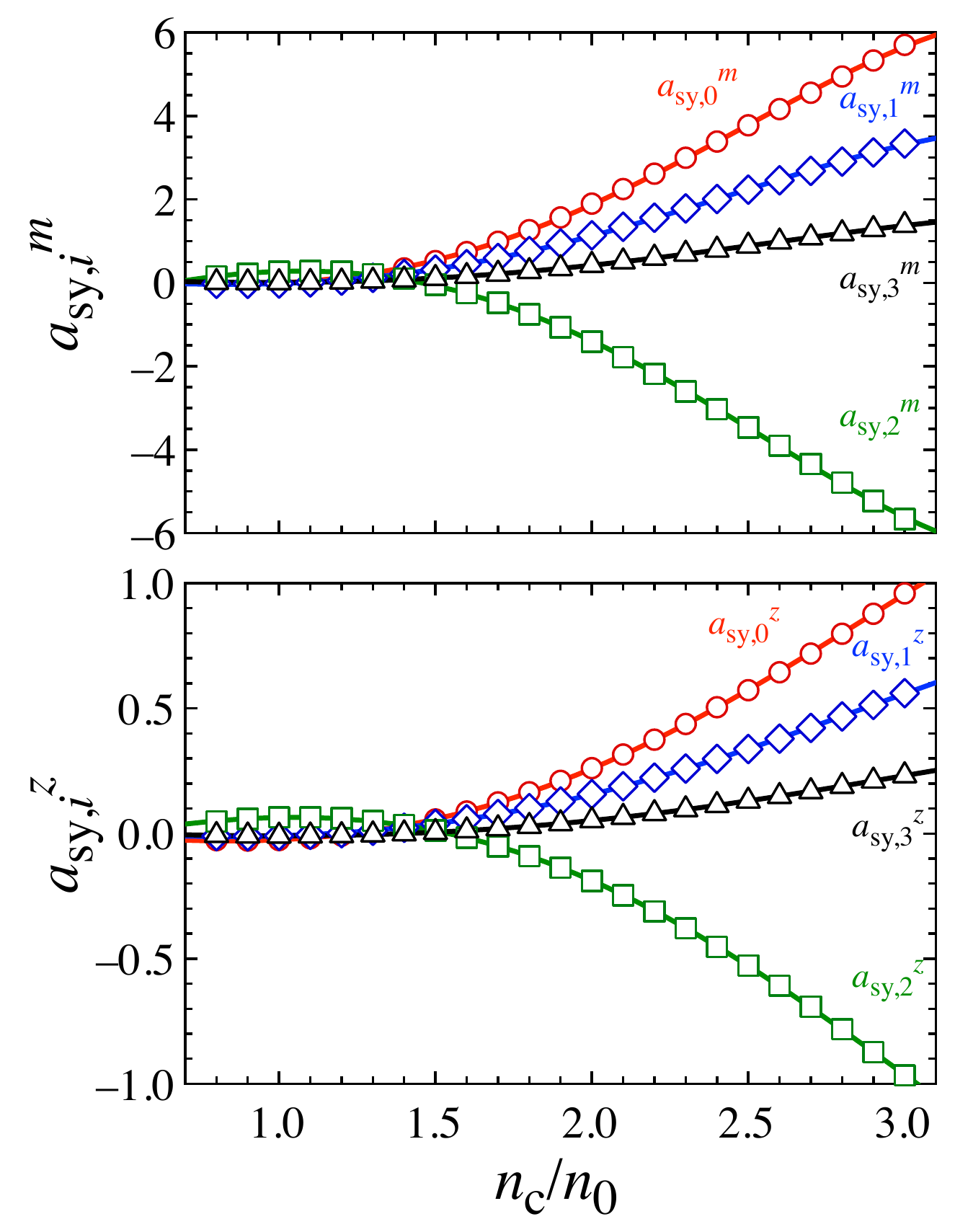} 
\end{center}
\caption{
The coefficients in Eqs. (\ref{eq:mmsy}) and (\ref{eq:zzsy}) are plotted as a function of ${\cal R}_c=n_c/n_0$, while the solid lines are the corresponding fitting given by Eqs. (\ref{eq:asym}) and (\ref{eq:asyz}).
}
\label{fig:asyimz}
\end{figure}

Up to now we consider to derive the empirical formulas with $\eta$, but another combination of the nuclear saturation parameters may be better to express the neutron star mass and redshift. Here, we consider to derive the empirical formulas as a function of $\eta_{sy}$ defined by Eq. (\ref{eq:etasy})
instead of $\eta$. In Fig. \ref{fig:Mzetasy} we plot the neutron star mass and redshift with $n_c/n_0=1, 2, 3$ constructed with each EOS, together with the fitting lines given by
\begin{gather}
  \frac{M_{\eta_{sy}}}{M_\odot} = a_{sy,0}^m + a_{sy,1}^m\ln(\eta_{sy,100}) + a_{sy,2}^m\eta_{sy,100} 
      + a_{sy,3}^m\eta_{sy,100}^2,  \label{eq:mmsy} \\
  z_{\eta_{sy}} = a_{sy,0}^z + a_{sy,1}^z\ln(\eta_{sy,100}) + a_{sy,2}^z\eta_{sy,100} 
      + a_{sy,3}^z\eta_{sy,100}^2, \label{eq:zzsy} 
\end{gather}
where $\eta_{sy,100}\equiv \eta_{sy}/(100\ {\rm MeV})$. Again, the coefficients, $a_{sy,i}^m$ and $a_{sy,i}^z$, depend on ${\cal R}_c$, which are shown in Fig. \ref{fig:asyimz}. In this figure, the marks denote numerical values determined by fitting with Eqs. (\ref{eq:mmsy}) and (\ref{eq:zzsy}) as in Fig. \ref{fig:Mzetasy}, while the solid lines denote the fitting of $a_{sy,i}^m$ and $a_{sy,i}^z$ as a function of ${\cal R}_c$ with
\begin{gather}
  a_{sy,i}^m({\cal R}_c) = a_{sy,i0}^m +a_{sy,i1}^m{\cal R}_c +a_{sy,i2}^m{\cal R}_c^3 
      +a_{sy,i3}^m {\cal R}_c^5, \label{eq:asym} \\
  a_{sy,i}^z({\cal R}_c) = a_{sy,i0}^z +a_{sy,i1}^z{\cal R}_c +a_{sy,i2}^z{\cal R}_c^3 
      +a_{sy,i3}^z {\cal R}_c^5, \label{eq:asyz} 
\end{gather}
where the concrete values of $a_{sy,ij}^m$ and $a_{sy,ij}^z$ for $i=0-3$ and $j=0-3$ are listed in Table \ref{tab:etasy}. Comparing Fig. \ref{fig:Mzetasy} to Fig. \ref{fig:Mzeta}, $\eta$ seems to be better than $\eta_{sy}$ in this stage, because a specific EOS model (e.g., OI 220) with $n_c/n_0=3$ largely deviates from the fitting line in Fig. \ref{fig:Mzetasy}.

\begin{table}
\caption{Values of $a_{sy,ij}^m$ and $a_{sy,ij}^z$ in Eqs. (\ref{eq:asym}) and (\ref{eq:asyz}).} 
\label{tab:etasy}
\begin {center}
\begin{tabular}{ccccc}
\hline\hline
 $j$  &  0 & 1 & 2  & 3     \\
\hline
 $a_{sy,0j}^m$ &  $0.6990$  & $-1.2408$  & $0.5655$  & $-0.02710$   \\
 $a_{sy,1j}^m$ &  $0.3220$  & $ -0.6504$  & $0.3266$  & $ -0.01596$   \\
 $a_{sy,2j}^m$ &  $-1.2165$  & $2.1443$  & $-0.6833$  & $0.03134$   \\
 $a_{sy,3j}^m$ &  $0.2314$  & $-0.3555$  & $0.1374$  & $-0.006172$   \\
 $a_{sy,0j}^z$  &  $0.05286$  & $-0.1492$  & $0.07355$  & $ -0.002614$   \\
 $a_{sy,1j}^z$  &  $0.03503$  & $-0.08519$  & $0.04252$  & $-0.001517$   \\
 $a_{sy,2j}^z$  &  $-0.1233$  & $0.2726$  & $-0.08809$  & $0.002965$   \\
 $a_{sy,3j}^z$  &  $0.02629$  & $-0.05430$  & $0.01929$  & $-0.0006297$   \\
\hline \hline
\end{tabular}
\end {center}
\end{table}

\begin{figure}[tbp]
\begin{center}
\includegraphics[scale=0.5]{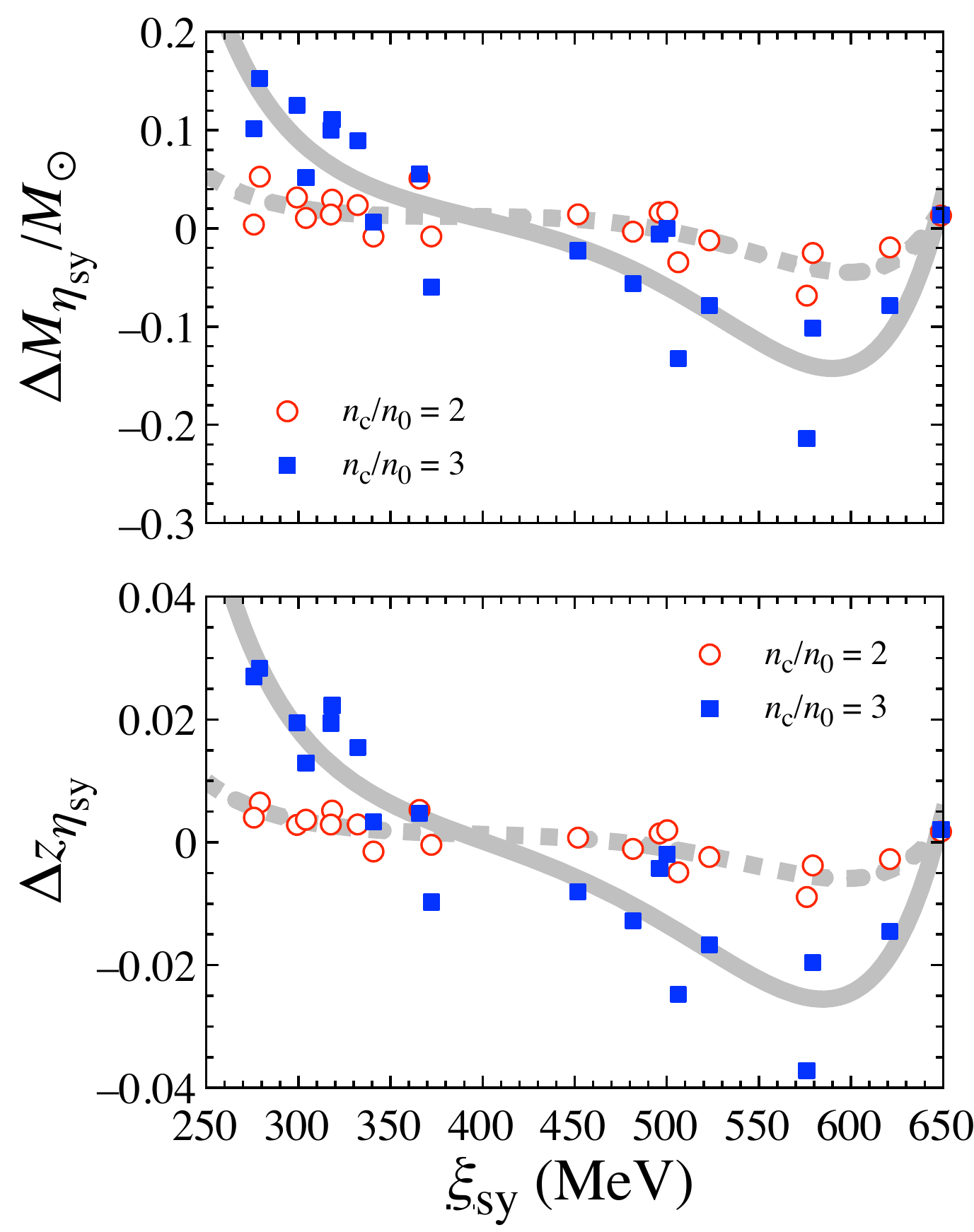} 
\end{center}
\caption{
$\Delta M_{\eta_{sy}}$ and $\Delta z_{\eta_{sy}}$ with $n_c/n_0=2$ and 3 are plotted as a function of $\xi_{sy}$, where the thick-solid lines for $n_c/n_0=3$ and thick-dotted lines for $n_c/n_0=2$ are fitting lines given by Eqs. (\ref{eq:dMxisy}) and (\ref{eq:dzxisy}).
}
\label{fig:dMdz-xisy}
\end{figure}

\begin{figure}[tbp]
\begin{center}
\includegraphics[scale=0.5]{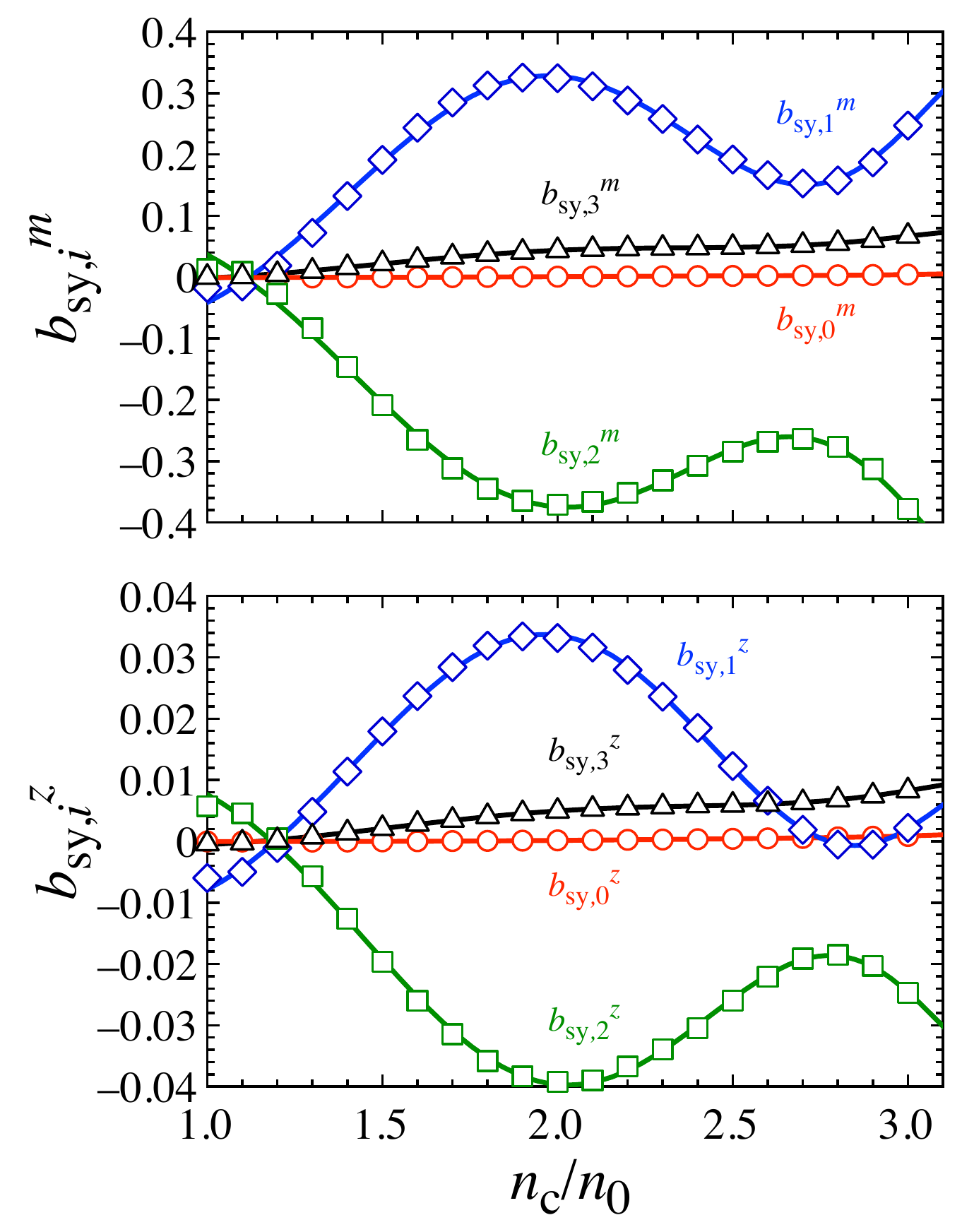} 
\end{center}
\caption{
The coefficients in the fitting formulas (Eqs. (\ref{eq:dMxisy}) and (\ref{eq:dzxisy})), $b_{sy,i}^m$ and $b_{sy,i}^z$ for $i=0-3$, are plotted as a function of $n_c/n_0$, while the solid lines are the fitting of those values as $n_c/n_0$ with the functional form given by Eqs. (\ref{eq:bsy0m}) -- (\ref{eq:bsyiz}).
}
\label{fig:bsyimz}
\end{figure}

Then, we consider to characterize the deviation of the neutron star mass and redshift estimated with the empirical formulas,  $M_{\eta_{sy}}({\cal R}_c,\eta_{sy})$ and $z_{\eta_{sy}}({\cal R}_c,\eta_{sy})$, given by Eqs. (\ref{eq:mmsy}) -- (\ref{eq:asyz}) from those determined as the TOV solution. Such a deviation is given by 
\begin{gather}
  \Delta M_{\eta_{sy}} = M_{\rm TOV} - M_{\eta_{sy}}({\cal R}_c,\eta_{sy}), \label{dMsy} \\
  \Delta z_{\eta_{sy}} = z_{\rm TOV} - z_{\eta_{sy}}({\cal R}_c,\eta_{sy}). \label{dzsy}   
\end{gather}
In the beginning we try to characterize $\Delta M_{\eta_{sy}}$ and $\Delta z_{\eta_{sy}}$ with a specific combination of $Q$ and $Q_{sym}$, because $K_{sym}$ is already included in the definition of $\eta_{sy}$, but eventually we find that the combination of $K_{sym}$, $Q$, and $Q_{sym}$ defined by Eq. (\ref{eq:xisy})
are suitable for this problem. In fact, as shown in Fig. \ref{fig:dMdz-xisy}, $\Delta M_{\eta_{sy}}$ and $\Delta z_{\eta_{sy}}$ are well fitted as a function of $\xi_{sy}$, where the open-circles (filled-squares) denote the values of $\Delta M_{\eta_{sy}}$ and $\Delta z_{\eta_{sy}}$ with $n_c/n_0=2$ ($n_c/n_0=3$), while the dotted (solid) lines denote the fitting of those values with  $n_c/n_0=2$ ($n_c/n_0=3$) by 
\begin{gather}
  \frac{\Delta M_{\eta_{sy}}}{M_\odot} = b_{sy,0}^m/\xi_{sy,500}^6+b_{sy,1}^m \xi_{sy,500}^7 
      + b_{sy,2}^m \xi_{sy,500}^8 + b_{sy,3}^m \xi_{sy,500}^{12},  \label{eq:dMxisy}  \\
  \Delta z_{\eta_{sy}} = b_{sy,0}^z/\xi_{sy,500}^6 +b_{sy,1}^z \xi_{sy,500}^7
      + b_{sy,2}^z \xi_{sy,500}^8 + b_{sy,3}^z \xi_{sy,500}^{12}.  \label{eq:dzxisy} 
\end{gather}
In these fitting formulas, $\xi_{sy,500}$ is defined as $\xi_{sy,500}\equiv \xi_{sy}/(500\ {\rm MeV})$, while $b_{sy,i}^m$ and $b_{sy,i}^z$ for $i=0-3$ are the adjusting coefficients, depending on $n_c/n_0$. In Fig. \ref{fig:bsyimz}, the values of $b_{sy,i}^m$ and $b_{sy,i}^z$ for $i=0-3$ are plotted as a function of $n_c/n_0$, where the solid lines are the fitting of those values with the functional form given by
\begin{gather}
  b_{sy,0}^m({\cal R}_c) = b_{sy,00}^m{\cal R}_c^2 +b_{sy,01}^m{\cal R}_c^4 +b_{sy,02}^m{\cal R}_c^6
      +b_{sy,03}^m {\cal R}_c^9, \label{eq:bsy0m} \\
  b_{sy,i}^m({\cal R}_c) = b_{sy,i0}^m{\cal R}_c^2 +b_{sy,i1}^m{\cal R}_c^4 +b_{sy,i2}^m{\cal R}_c^5
      +b_{sy,i3}^m {\cal R}_c^7+b_{sy,i4}^m {\cal R}_c^9, \label{eq:bsyim} \\
  b_{sy,0}^z({\cal R}_c) = b_{sy,00}^z{\cal R}_c^3 +b_{sy,01}^z{\cal R}_c^4 +b_{sy,02}^z{\cal R}_c^8
      +b_{sy,03}^z {\cal R}_c^9, \label{eq:bsy0z} \\
  b_{sy,i}^z({\cal R}_c) = b_{sy,i0}^z{\cal R}_c^2 +b_{sy,i1}^z{\cal R}_c^4 +b_{sy,i2}^z{\cal R}_c^5
      +b_{sy,i3}^z {\cal R}_c^7+b_{sy,i4}^z {\cal R}_c^9. \label{eq:bsyiz}
\end{gather}
The coefficients in these equations, $b_{sy,ij}^m$ and $b_{sy,ij}^z$, are concretely listed in Table \ref{tab:bsym}.

\begin{table*}
\caption{Values of $b_{sy,ij}^m$ and $b_{sy,ij}^z$ in Eqs. (\ref{eq:bsy0m}) -- (\ref{eq:bsyiz}).} 
\label{tab:bsym}
\begin {center}
\begin{tabular}{cccccc}
\hline\hline
 $j$  &  0 & 1 & 2  & 3  & 4   \\
\hline
 $b_{sy,0j}^m$ &  $-9.146\times 10^{-4}$  & $4.823\times 10^{-4}$  & $-5.734\times 10^{-5}$  & $7.712\times 10^{-7}$ & ---  \\
 $b_{sy,1j}^m$ &  $-3.719\times 10^{-1}$  & $6.652\times 10^{-1}$  & $-3.572\times 10^{-1}$  & $2.324\times 10^{-2}$  &  $-7.279\times 10^{-4}$ \\
 $b_{sy,2j}^m$ &  $3.684\times 10^{-1}$  & $-6.647\times 10^{-1}$  & $3.551\times 10^{-1}$  & $-2.302\times 10^{-2}$  & $7.217\times 10^{-4}$ \\
 $b_{sy,3j}^m$ &  $-2.956\times 10^{-2}$  & $5.484\times 10^{-2}$  & $-2.879\times 10^{-2}$  & $1.839\times 10^{-3}$ & $-5.768\times 10^{-5}$  \\
 $b_{sy,0j}^z$  &  $-1.032\times 10^{-4}$  & $7.160\times 10^{-5}$  & $-1.388\times 10^{-6}$  & $3.541\times 10^{-7}$ & ---  \\
 $b_{sy,1j}^z$  &  $-4.730\times 10^{-2}$  & $7.798\times 10^{-2}$  & $-4.105\times 10^{-2}$  & $2.563\times 10^{-3}$ & $-7.723\times 10^{-5}$ \\
 $b_{sy,2j}^z$  &  $4.726\times 10^{-2}$  & $-7.785\times 10^{-2}$  & $4.065\times 10^{-2}$  & $-2.522\times 10^{-3}$ & $7.586\times 10^{-5}$ \\
 $b_{sy,3j}^z$  &  $-3.880\times 10^{-3}$  & $6.386\times 10^{-3}$  & $-3.246\times 10^{-3}$  & $1.967\times 10^{-4}$ & $-5.867\times 10^{-6}$ \\
\hline \hline
\end{tabular}
\end {center}
\end{table*}

\begin{figure*}[tbp]
\begin{center}
\includegraphics[scale=0.5]{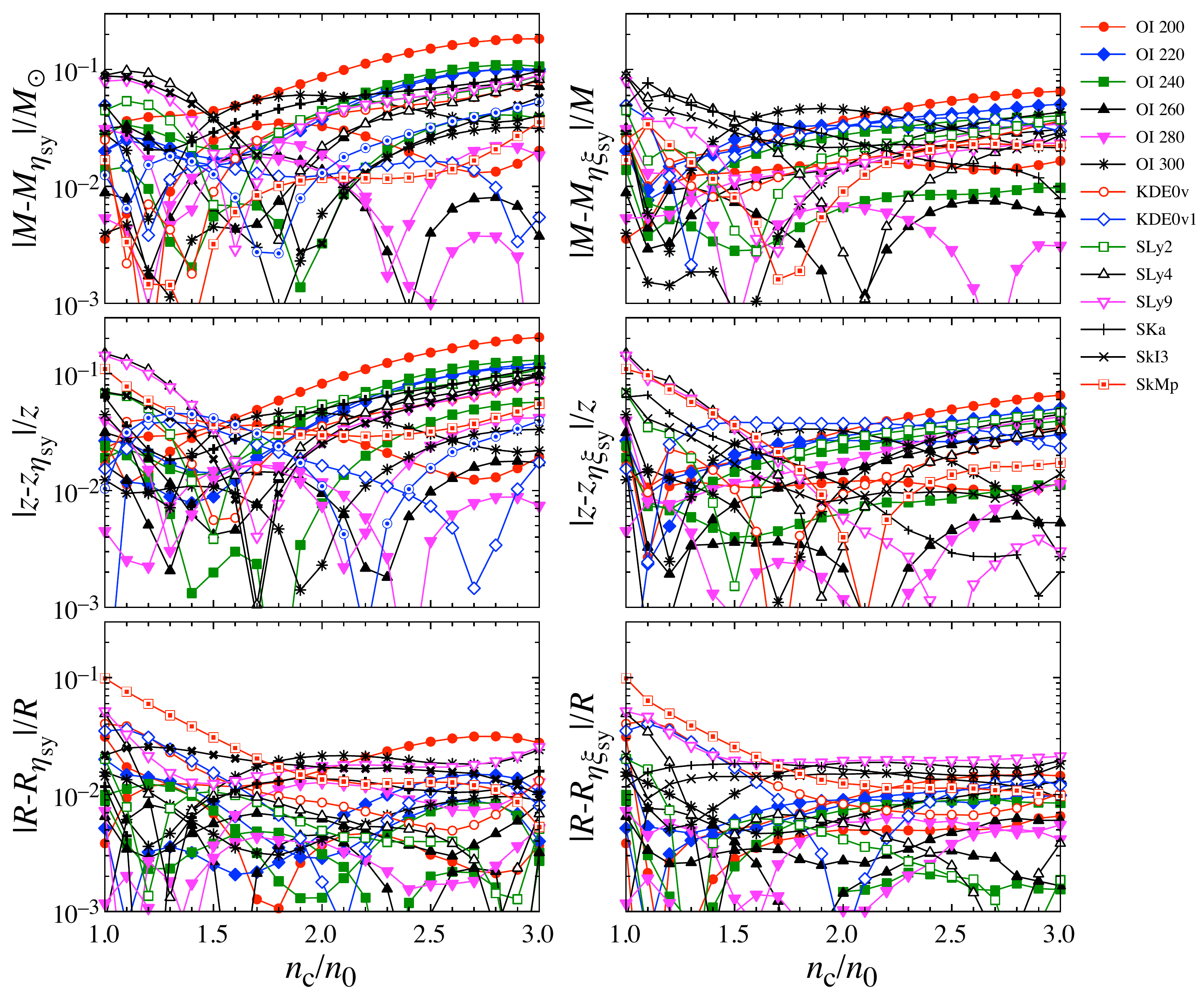} 
\end{center}
\caption{
Same as Fig. \ref{fig:dMzR}, but with the empirical formulas, $M_{\eta_{sy}}({\cal R}_c,\eta_{sy})$, $z_{\eta_{sy}}({\cal R}_c,\eta_{sy})$, $M_{\eta\xi_{sy}}({\cal R}_c,\eta_{sy},\xi_{sy})$, and $z_{\eta\xi_{sy}}({\cal R}_c,\eta_{sy},\xi_{sy})$.
}
\label{fig:dMzRsy}
\end{figure*}

Now, we get the alternative empirical formulas for the neutron star mass and gravitational redshift as a function of ${\cal R}_c$, $\eta_{sy}$, and $\xi_{sy}$:
\begin{gather}
  \frac{M_{\eta\xi_{sy}}}{M_\odot} = \frac{M_{\eta_{sy}}({\cal R}_c,\eta_{sy})}{M_\odot} 
       + \frac{\Delta M_{\eta_{sy}}({\cal R}_c,\xi_{sy})}{M_\odot}, \label{eq:Mxisy} \\ 
  z_{\eta\xi_{sy}} = z_{\eta_{sy}}({\cal R}_c,\eta_{sy}) + \Delta z_{\eta_{sy}}({\cal R}_c,\xi_{sy}), \label{eq:zxisy}
\end{gather}
where the first terms are given by Eqs. (\ref{eq:mmsy}) -- (\ref{eq:asyz}) and the second terms are given by Eqs. (\ref{eq:dMxisy}) -- (\ref{eq:bsyiz}). In order to check how well one can estimate the neutron star mass and gravitational redshift with the empirical formulas with $\eta_{sy}$, i.e., $M_{\eta_{sy}}({\cal R}_c,\eta_{sy})$, $z_{\eta_{sy}}({\cal R}_c,\eta_{sy})$, $M_{\eta\xi_{sy}}({\cal R}_c,\eta_{sy},\xi_{sy})$, and $z_{\eta\xi_{sy}}({\cal R}_c,\eta_{sy},\xi_{sy})$, we calculate the relative deviation from the TOV solutions constructed with concrete EOSs and show the absolute value of it in Fig. \ref{fig:dMzRsy}, where the top and middle panels correspond to the mass and gravitational redshift, while the bottom panels are the relative deviation of the radius estimated with the empirical formulas for mass and gravitational redshift. Comparing to Fig. \ref{fig:dMzR}, one can see that the empirical formulas with $\eta_{sy}$ are the same level as or better than those with $\eta$. In fact, with respect to the canonical neutron star models, one can estimate the mass (radius) within $\sim 7\%$ ($\sim 2\%$) accuracy, using the empirical relations, $M_{\eta\xi_{sy}}({\cal R}_c,\eta_{sy},\xi_{sy})$ and $z_{\eta\xi_{sy}}({\cal R}_c,\eta_{sy},\xi_{sy})$. 
We note that one can accurately estimate the radius by using the empirical formulas for the mass and gravitational redshift again, even though the dependence of the mass and gravitational redshift on $\eta_{sy}$ are quite similar, as shown in Fig. \ref{fig:Mzetasy}.

\begin{table}
\caption{Correspondence between the empirical formulas and their equations.} 
\label{tab:formulas}
\begin {center}
\begin{tabular}{ll}
\hline\hline
 empirical forumula  & corresponding equations     \\
\hline
 $M_\eta({\cal R}_c,\eta)$      &  (\ref{eq:mm}) and (\ref{eq:am})    \\
 $M_{\eta\xi}({\cal R}_c,\eta,\xi)$ &   (\ref{eq:Mxi}) with (\ref{eq:mm}), (\ref{eq:am}), (\ref{eq:dMxi}), \& (\ref{eq:bm}) \\
 $M_{\eta_{sy}}({\cal R}_c,\eta_{sy})$                &  (\ref{eq:mmsy}) and (\ref{eq:asym})   \\
 $M_{\eta\xi_{sy}}({\cal R}_c,\eta_{sy},\xi_{sy})$ &  (\ref{eq:Mxisy}) with (\ref{eq:mmsy}), (\ref{eq:asym}), (\ref{eq:dMxisy}), (\ref{eq:bsy0m}), \& (\ref{eq:bsyim})  \\
 $z_\eta({\cal R}_c,\eta)$      &   (\ref{eq:zz}) and (\ref{eq:az})  \\
 $z_{\eta\xi}({\cal R}_c,\eta,\xi)$ &  (\ref{eq:zxi}) with (\ref{eq:zz}), (\ref{eq:az}), (\ref{eq:dzxi}), \& (\ref{eq:bz})   \\
 $z_{\eta_{sy}}({\cal R}_c,\eta_{sy})$                 &   (\ref{eq:zzsy}) and (\ref{eq:asyz}) \\
 $z_{\eta\xi_{sy}}({\cal R}_c,\eta_{sy},\xi_{sy})$ &   (\ref{eq:zxisy}) with (\ref{eq:zzsy}), (\ref{eq:asyz}), (\ref{eq:dzxisy}), (\ref{eq:bsy0z}), \& (\ref{eq:bsyiz})   \\
\hline \hline
\end{tabular}
\end {center}
\end{table}

\begin{figure}[tbp]
\begin{center}
\includegraphics[scale=0.6]{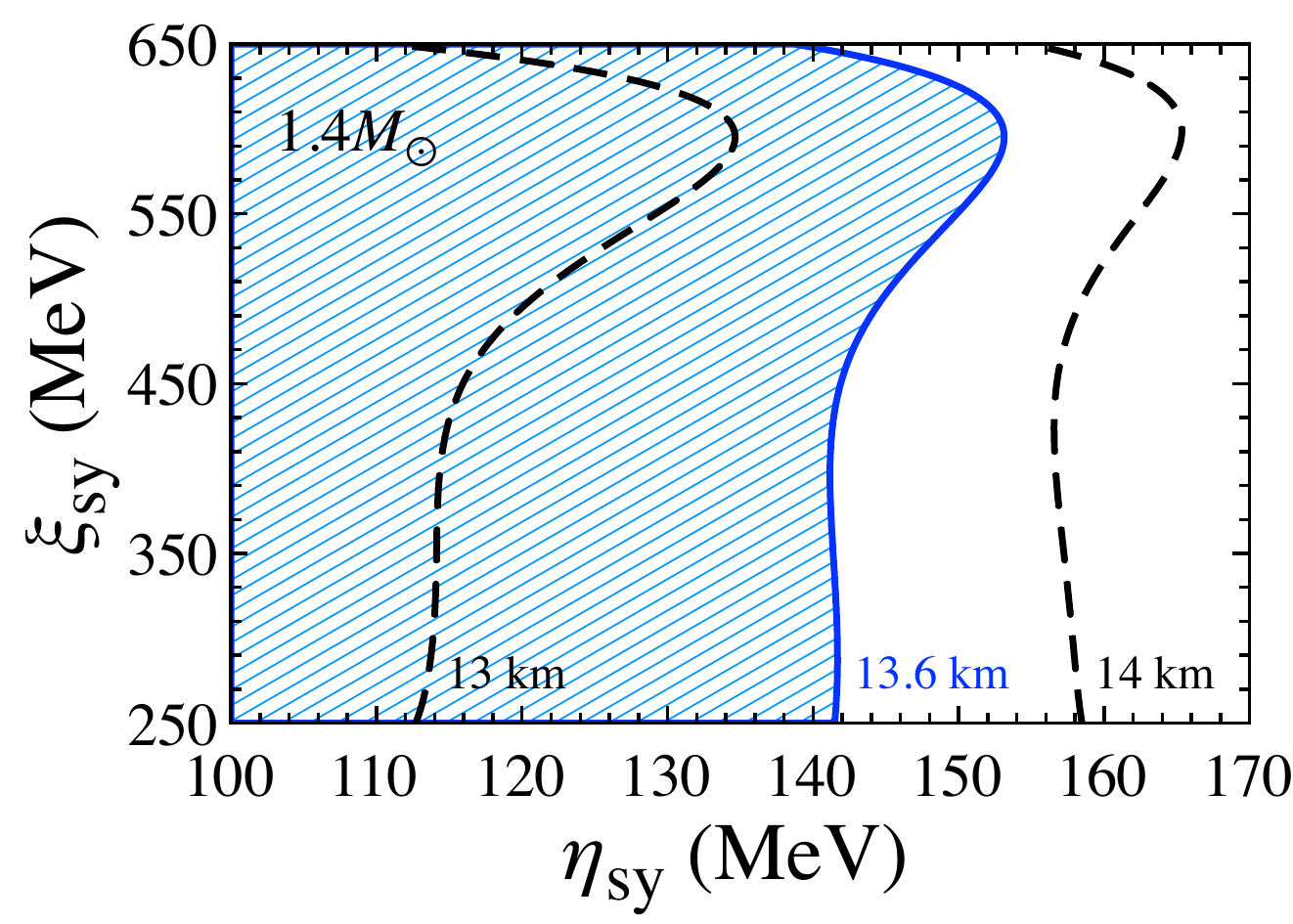} 
\end{center}
\caption{
Constraint on the parameter space with $\eta_{sy}$ and $\xi_{sy}$ obtained from the constraint on the $1.4M_\odot$ neutron star radius, $R_{1.4}$, with the gravitational wave event, GW1708107, i.e., $R_{1.4}\le 13.6$ km \cite{Annala18}. In this figure, the shaded region corresponds to the allowed region.
}
\label{fig:const}
\end{figure}

Finally, in Table \ref{tab:formulas}, we show the corresponding equations for the empirical relations obtained in this study. In addition, using the estimation of the neutron star mass and radius with the empirical formulas, $M_{\eta\xi_{sy}}({\cal R}_c,\eta_{sy},\xi_{sy})$ and $z_{\eta\xi_{sy}}({\cal R}_c,\eta_{sy},\xi_{sy})$, we make a constraint on the parameter space with $\eta_{sy}$ and $\xi_{sy}$. That is, owing to the gravitational wave observations at the GW170817, the tidal deformability of neutron star has been constrained, which tells us that the $1.4M_\odot$ neutron star radius should be less than 13.6 km \cite{Annala18}. In practice, assuming that $M=1.4M_\odot$ together with $M_{\eta\xi_{sy}}({\cal R}_c,\eta_{sy},\xi_{sy})$ and $z_{\eta\xi_{sy}}({\cal R}_c,\eta_{sy},\xi_{sy})$, one can estimate the stellar radius with the given values of $\eta_{sy}$ and $\xi_{sy}$. In Fig. \ref{fig:const} we plot the combination of $\eta_{sy}$ and $\xi_{sy}$ so that the radius becomes 13 and 14 km with dashed lines and 13.6 km with solid line. So, one can see that  the shaded region corresponds to the allowed region, considering the constraint through the GW170817.

\section{Conclusion}
\label{sec:Conclusion}

The neutron star mass and radius are one of the most important observables to constrain the EOS for dense matter. In fact, some of astronomical observations could make a constraint on EOS, essentially for higher density region. On the other hand, the terrestrial nuclear experiments constrain the nuclear properties especially around the nuclear saturation density, which enables us to screen the EOSs. So, at least the neutron star models for lower density region are strongly associated with the nuclear saturation parameters. In this study, we propose the empirical formulas for the neutron star mass and gravitational redshift as a function of the central density and the suitable combination of nuclear saturation parameters, which are applicable 
to the stellar models constructed with the central density up to threefold nuclear density. Combining both empirical relations, the stellar radius is also estimated. Our empirical formulas can directly connect the neutron star properties to the nuclear saturation parameters, which helps us to imagine the neutron star mass and radius with the specific values of saturation parameters constrained via experiments, and vice versa. As an application with our empirical formulas, we constrain the parameter space of the nuclear saturation parameters, considering the constraint on the neutron star radius through the gravitational wave observations at the GW170817. Although the current constraint is still poor, one can discuss the nuclear saturation parameters more severely, as the astronomical observations would increase.
In this study, we focus only on the empirical relations for the neutron star mass and gravitational redshift, but it must be also possible to derive the empirical formulas for the other neutron star bulk properties, such as the moment of inertia or Love number, as in Ref. \cite{SSB16}. We will consider these topics somewhere in the future.

\acknowledgments

This work is supported in part by Japan Society for the Promotion of Science (JSPS) KAKENHI Grant Numbers 
JP18K13551,  
JP19KK0354,  
JP20H04753,  and 
JP21H01088,  
and by Pioneering Program of RIKEN for Evolution of Matter in the Universe (r-EMU).



\begin{thebibliography}{999}


\bibitem{D10} 
   P. Demorest, T. Pennucci, S. Ransom, M. Roberts, and J. Hessels, Nature {\bf 467}, 1081 (2010).

\bibitem{A13} 
   J. Antoniadis {\it et al.}, Science {\bf 340}, 6131 (2013).

\bibitem{C20}    
   H. T. Cromartie {\it et al.}, Nature Astronomy {\bf 4}, 72 (2020).

\bibitem{F21}    
   E. Fonseca et al., Astrophys. J. 915, L12 (2021).

\bibitem{PFC83} 
   K. R. Pechenick, C. Ftaclas, and J. M. Cohen, Astrophys. J. {\bf 274}, 846 (1983).

\bibitem{LL95} 
   D. A. Leahy and L. Li, Mon. Not. R. Astron. Soc. {\bf 277}, 1177 (1995).

\bibitem{PG03} 
   J. Poutanen and M. Gierlinski, Mon. Not. R. Astron. Soc. {\bf 343}, 1301 (2003).

\bibitem{PO14} 
   D. Psaltis and F. \"{O}zel, Astrophys. J. {\bf 792}, 87 (2014). 

\bibitem{SM18} 
   H. Sotani and U. Miyamoto, Phys. Rev. D {\bf 98}, 044017 (2018); {\bf 98}, 103019 (2018).

\bibitem{Sotani20a} 
   H. Sotani, Phys. Rev. D {\bf 101}, 063013 (2020).

\bibitem{Riley19}  
   T. E. Riley {\it et al.}, Astrophys. J.  {\bf 887}, L21 (2019).
   
\bibitem{Miller19} 
   M. C. Miller {\it et al.}, Astrophys. J.  {\bf 887}, L24 (2019).

\bibitem{Riley21} 
   T. E. Riley et al., Astrophys. J. {\bf 918}, L27 (2021).
   
\bibitem{Miller21} 
   M. C. Miller et al., Astrophys. J. {\bf 918}, L28 (2021).

\bibitem{gw170817} 
   B. P. Abbott et al. (The LIGO Scientific Collaboration and the Virgo Collaboration), Phys. Rev. Lett. {\bf 119}, 161101 (2017).

\bibitem{Annala18}  
   E. Annala, T. Gorda, A. Kurkela, and A. Vuorinen, Phys. Rev. Lett. {\bf 120}, 172703 (2018).

\bibitem{AK1996}
   N. Andersson and K. D. Kokkotas, Phys.\ Rev.\ Lett.\ {\bf 77}, 4134 (1996).

\bibitem{AK1998}
   N. Andersson and K. D. Kokkotas, Mon.\ Not.\ R. Astron.\ Soc.\ {\bf 299}, 1059 (1998).

\bibitem{STM2001}
   H. Sotani, K. Tominaga, and K. I. Maeda, Phys.\ Rev.\ D {\bf 65}, 024010 (2001).

\bibitem{SH2003}
   H. Sotani and T. Harada, Phys.\ Rev.\ D {\bf 68}, 024019 (2003);
   H. Sotani, K. Kohri, and T. Harada, {\it ibid}.\ {\bf 69}, 084008 (2004).

\bibitem{SYMT2011}
   H. Sotani, N. Yasutake, T. Maruyama, and T. Tatsumi, Phys.\ Rev.\ D {\bf 83} 024014 (2011).

\bibitem{PA2012}
   A. Passamonti and N. Andersson, Mon.\ Not.\ R. Astron.\ Soc.\ {\bf 419}, 638 (2012).

\bibitem{DGKK2013}
   D. D. Doneva, E. Gaertig, K. D. Kokkotas, and C. Kr\"{u}ger, Phys.\ Rev.\ D {\bf 88}, 044052 (2013).

\bibitem{Sotani2020}
   H. Sotani, Phys.\ Rev.\ D {\bf 102}, 063023 (2020); 103021 (2020); {\bf 103}, 123015 (2021).

\bibitem{SD2021}
   H. Sotani and A. Dohi, accepted in PRD.

\bibitem{SPIRIT}
   J. Estee et al. (S$\pi$RIT), Phys. Rev. Lett. 126, 162701 (2021).

\bibitem{PREXII}
   B. T. Reed, F. J. Fattoyev, C. J. Horowitz, and J. Piekarewicz, Phys. Rev. Lett. 126, 172503 (2021).

\bibitem{SIOO14} 
   H. Sotani, K. Iida, K. Oyamatsu, and A. Ohnishi, Prog. Theor. Exp. Phys. {\bf 2014}, 051E01 (2014).

\bibitem{OI03}  
   K. Oyamatsu and K. Iida, Prog. Theor. Phys. {\bf 109}, 631 (2003).

\bibitem{OI07}  
   K. Oyamatsu and K. Iida, Phys. Rev. C {\bf 75}, 015801 (2007).




\bibitem{KDE0v}  
   B. K. Agrawal, S. Shlomo and V. Kim Au, Phys. Rev. C {\bf 72}, 014310 (2005).

\bibitem{SLy4}  
   F. Douchin and P. Haensel, Astron. Astrophys. {\bf 380}, 151 (2001).

\bibitem{SLy9}  
   E. Chabanat, {\it Interactions effectives pour des conditions extremes d'isospin}, Ph.D. thesis, University Claude Bernard Lyon-I (1995).

\bibitem{SKa}  
   H. S. K\"{o}hler, Nucl. Phys. A {\bf 258}, 301 (1976).

\bibitem{SkI3}  
   P. -G. Reinhard and H. Flocard, Nucl. Phys. A {\bf 584}, 467 (1995).

\bibitem{SkMp}  
   L. Bennour, P-H. Heenen, P. Bonche, J. Dobaczewski, and H. Flocard, Phys. Rev. C {\bf 40}, 2834 (1989).
   
\bibitem{Shen}  
   H. Shen, H. Toki, K. Oyamatsu, and K. Sumiyoshi, Nucl. Phys. {\bf A637}, 435 (1998).

\bibitem{Togashi17}  
   H. Togashi, K. Nakazato, Y. Takehara, S. Yamamuro, H. Suzuki, and M. Takano, Nucl. Phys. A {\bf 961}, 78 (2017).
  
   

\bibitem{OHKT17}
   M. Oertel, M. Hempel, T. Kl\"{a}hn, and S. Typel, Rev. Mod. Phys. {\bf 89}, 015007 (2017).

\bibitem{Li19}
   B. A. Li, P. G. Krastev, D. H. Wen, and N. B. Zhang, Euro. Phys. J. A {\bf 55}, 117 (2019).

\bibitem{KM13}
   E. Khan and J. Margueron, Phys. Rev. C {\bf 88}, 034319 (2013).

\bibitem{LL13}
   J. M. Lattimer and Y. Lim, Astrophys. J. {\bf 771}, 51 (2013).



\bibitem{Danielewicz09}
   P. Danielewicz and J. Lee, Nucl. Phys. A {\bf 818},36 (2009).

\bibitem{Tews17}
   I. Tews, J. M. Lattimer, A. Ohnishi, and E. E. Kolomeitsev, Astrophys. J. {\bf 848}, 105 (2017).



\bibitem{SNN22}
  H. Sotani, N. Nishimura, and T. Naito, arXiv:2203.05410 [nucl-th]. 

\bibitem{SSB16} 
   H. O. Silva, H. Sotani, and E. Berti, Mon. Not. R. Astron. Soc. {\bf 459}, 4378 (2016).   

\bibitem{Sotani17}  
   H. Sotani, Phys. Rev. C {\bf 95}, 025802 (2017).

\bibitem{SK17}  
   H. Sotani and K. D. Kokkotas, Phys. Rev. D {\bf 95}, 044032 (2017).




\end{thebibliography}

\end{document}